\documentclass[twocolumn,twocolappendix,trackchanges]{aastex63}
\usepackage{amsmath}
\usepackage{amssymb}
\usepackage{bm}
\usepackage{graphicx}
\usepackage{color}
\usepackage{float}
\usepackage{multirow}
\usepackage{CJK}
\usepackage[T1]{fontenc}
\usepackage[hang,flushmargin]{footmisc}

\definecolor{darkgreen}{rgb}{0,0.35,0}
\definecolor{blue}{rgb}{0,0,1}

\newcommand{\be}{\begin{eqnarray}}
\newcommand{\ee}{\end{eqnarray}}

\begin{document}

\begin{CJK*}{UTF8}{ipxm}
\title{Multi-wavelength Constraints on Dust Dynamics and Size Evolution in Protoplanetary Disk Rings. I. Method}
\author[0009-0002-8049-6554]{Linhan Yang\begin{CJK*}{UTF8}{gbsn}(杨林翰)\end{CJK*}}
\affiliation{Shanghai Astronomical Observatory, Chinese Academy of Sciences, Shanghai 200030, China}
\affiliation{School of Astronomy and Space Science, University of Chinese Academy of Sciences, Beijing 100049, China}

\author[0000-0002-7329-9344]{Ya-Ping Li\begin{CJK*}{UTF8}{gbsn}(李亚平)\end{CJK*}}
\affiliation{Shanghai Astronomical Observatory, Chinese Academy of Sciences, Shanghai 200030, China}

\author[0000-0001-9290-7846]{Ruobing Dong\begin{CJK*}{UTF8}{gbsn}(董若冰)\end{CJK*}}
\affiliation{Kavli Institute for Astronomy and Astrophysics, Peking University, Beijing 100871, China}

\author[0000-0003-1958-6673]{Kiyoaki Doi (土井聖明)}
\affiliation{Max-Planck Institute for Astronomy, Königstuhl 17, D-69117 Heidelberg, Germany}

\author[0000-0003-2300-2626]{Hauyu Baobab Liu(呂浩宇)}
\affiliation{Department of Physics, National Sun Yat-Sen University, No. 70, Lien-Hai Road, Kaohsiung City 80424, Taiwan, R.O.C.}
\affiliation{Center of Astronomy and Gravitation, National Taiwan Normal University, Taipei 116, Taiwan}

\author[0000-0002-7575-3176]{Pinghui Huang\begin{CJK*}{UTF8}{gbsn}(黄平辉)\end{CJK*}}
\affiliation{CAS Key Laboratory of Planetary Sciences, Purple Mountain Observatory, Chinese Academy of Sciences, Nanjing 210008, China}
\affiliation{Department of Physics \& Astronomy, University of Victoria, Victoria, British Columbia, V8P 5C2, Canada}
\affiliation{Institute for Advanced Study, Tsinghua University, Beijing 100084, China}
\correspondingauthor{Ya-Ping Li, Ruobing Dong}
\email{liyp@shao.ac.cn, rbdong@pku.edu.cn}

\begin{abstract}
Observations with the Atacama Large Millimeter/submillimeter Array (ALMA) and the Jansky Very Large Array (JVLA) have revealed many dust rings in protoplanetary disks, often interpreted as dust traps at gas pressure bumps. Previous studies have typically modeled these rings by assuming a single dust species in drift-diffusion equilibrium, neglecting dust size evolution resulting from coagulation and fragmentation. In this work, we perform numerical simulations that incorporate both dust-gas dynamics (drift and diffusion) and dust size evolution. Our results show that the radial distributions of different dust species (up to the fragmentation limit) are nearly identical in the dust ring, as dust growth dominates over drift and diffusion (e.g., with a typical dust-to-gas ratio of $\epsilon \sim 10^{-2}$). Building on this finding, we develop a comprehensive, self-consistent analytical theory that describes the dust ring structure while explicitly accounting for size evolution effects. Our model provides a unified framework for interpreting multi-wavelength observations by linking the physical dust distribution to the observed ring properties, thus laying the foundation for future observational modeling.

\end{abstract}



\section{Introduction}
Planets are thought to form and evolve within the protoplanetary disks (PPDs). These disks are primarily composed of gas, with a small fraction of solid material in the form of dust, which constitutes the building blocks of planets \citep{Testi_2014, Miotello_2023}. The Atacama Large Millimeter/sub-millimeter Array (ALMA), with its long-baseline configuration, has enabled high-resolution observations of these disks, revealing the widespread presence of dust substructures \citep{ALMA_2015}.

A prominent substructure observed in these disks is the concentric ring formation \citep{Andrews_2020, Bae_2023, Long_2018, Andrews_2018, Sierra2021}. These rings, which are axisymmetric features in millimeter continuum maps, typically exhibit radial intensity profiles that can be approximated by a Gaussian distribution \citep{Dullemond_2018, rosotti_2020, Carvalho2024}:

\begin{equation}
    I\mathrm{_{ring}}(r)=I_0 \exp{\left(-\frac{(r-r_0)^2}{2w_\mathrm{i}^2}\right)},
    \label{eq:I_ring}
\end{equation}
where the standard deviation $w\mathrm{_{i}}$ 
is defined in this study as the observed width of the ring.

Understanding these ringed substructures is of great significance, as they provide valuable insights into the physical processes governing dust evolution and gas-dust interactions within PPDs, thereby setting the initial stage for planet formation \citep{Rosotti_2023,Jiang_2023}.

The prevailing theory attributes the observed rings to dust trapping, with the spatial distribution of grains evolving passively in response to gas dynamics \citep[for a recent review, see][]{Bae_2023}. Other scenarios include the snow line \citep{Zhang2015, Okuzumi2016, Pinilla_2017}, dust-induced instabilities \citep{Gonzalez_2017} and traffic jam effects \citep{Jiang_2021,Carrera_2021}.

This work focus on the dust-trapping model. In this scenario, a gas pressure bump, caused by either a temperature or surface density perturbation, is required, though the former tends to be unstable in the absence of a surface density component \citep{Eonho_Chang2023}. The latter can result from density perturbations induced by embedded planets (e.g., \citealt{Lin_1986, Bae_2017, Dong_2015a, Dong_2017, Li2019b, Meru2019, Chen2021}, see \citealt{Paardekooper2023} for a review) or other dynamical processes, such as magnetohydrodynamics zonal flows \citep{Bai_Stone_2014, Johansen2009, Krapp2018, Cui2021}, dead zones \citep{Lyra2008, Li2019}, magnetized disk winds \citep{Suriano2017, Riols2019}, vertical shear instability \citep{Nelson2013, Lin2015}, the eccentric mode \citep{Lubow1991, LiJ2021}, vortex-disk interactions \citep{Ma2025}, and secular gravitational instability \citep{Youdin2011, Tominaga2019}. 

Within the pressure bump, drag forces arising from the velocity difference between dust and gas cause dust grains to drift toward the pressure maximum. Meanwhile, gas turbulence induces dust diffusion, broadening the ring. The drift-diffusion equilibrium for a single dust species (Eq.~\ref{eq:Dullemond}) is widely used to interpret ring widths measured at a single wavelength \citep{rosotti_2020, Carvalho2024, Dullemond_2018}.

However, previous studies have largely neglected dust collisions, induced by gas turbulence, Brownian motion, differential settling, and other physical processes. These collisions play a critical role in dust evolution, leading to coagulation, fragmentation, erosion, and bouncing of dust aggregates \citep[see][for a recent review]{Birnstiel_2024}. Moreover, in typical PPDs, dust collisions operate on the shortest timescales and dominate dust evolution.

With the advent of multi-wavelength, high-resolution observations of PPDs \citep[e.g.,][]{Ueda_2020,Carrasco-Gonz_2019,Sierra2021,Guidi_2022,Tsukagoshi2022,Han2023,Yang2023,Carvalho2024}, the brightness profiles of dust rings have been observed to vary across wavelengths\citep[e.g.,][]{Doi_2023,Carvalho2024,Sierra_2025}. 
While the spectral energy distribution (SED) has been used to infer dust size distributions \citep[e.g.,][]{Birnstiel_2018,LiD2024}, which inherently reflect the outcomes of dust collisions, interpretations of ring width variations across multiple wavelengths have largely relied on the drift-diffusion equilibrium scenario \citep[e.g.,][]{Doi_2023}, which neglects dust collisions. A comprehensive understanding of both spectral and spatial features requires integrating these approaches into a unified framework that self-consistently accounts for multi-species dust dynamics, including coagulation, fragmentation, diffusion, and drift.

This integration presents a significant challenge, as it requires solving integro-differential equations governing dust evolution, typically through computationally intensive numerical simulations. Several studies have employed hydrodynamical models to investigate the coupled evolution of dust and gas, incorporating coagulation and fragmentation in PPDs \citep{Li2019,Drazkowska2019,Stammler_2019,Laune_2020,Li2020,Chen2020,Ho2024}. However, the computational cost of such simulations scales exponentially with the number of free parameters, severely constraining the feasibility of extensive parameter studies and rendering exhaustive exploration impractical.

To enable more efficient modeling of steady-state dust rings and to provide a robust tool to constrain dust properties from multi-wavelength observations of rings, we introduce a novel, fully self-consistent analytical framework for dust rings incorporating dust drift, diffusion, coagulation, and fragmentation. This theory is calibrated against numerical simulations.

In a subsequent paper, we will apply this model to multi-wavelength observations of PPDs to constrain both dust and gas properties.

The structure of this paper is as follows: Section~\ref{sec:phys} outlines the key dust physics. Sections~\ref{sec:sim_setup} and \ref{sec:sim_result} present the numerical simulation setups and results, respectively. Section~\ref{sec:analytical_fits} develops the analytical framework for modeling dust ring dynamics, which could be applied in future observational modeling. Finally, Section~\ref{sec:summary} summarizes our findings.

\section{Dust Physics}\label{sec:phys}
\begin{figure}
\includegraphics[width=\columnwidth]{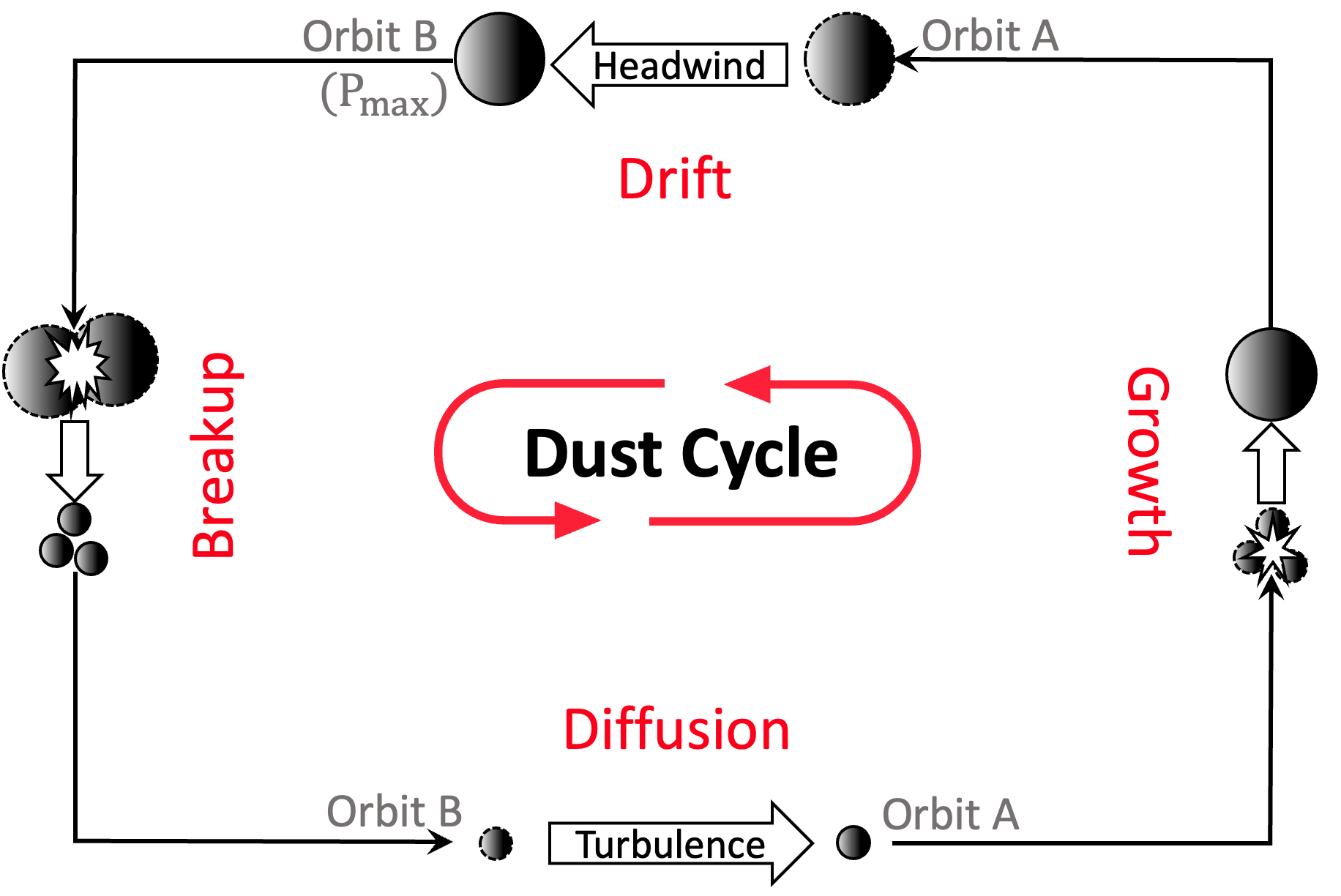}
\caption{Illustration of the primary physics around the pressure bump. Large pebbles, which drift efficiently, become trapped at the pressure maximum(upper), while small particles diffuse effectively and escape the trap(lower). Dust collisions facilitate interactions between these two populations. Large pebbles can fragment into smaller grains and diffuse away(left), while small grains can coagulate into larger pebbles and drift inward(right). This dynamics, identified in numerical simulations(Figure~\ref{fig:fiducial}), leads to both small and large grains share a similar spatial distribution when they are in a coagulation-fragmentation equilibrium, as further illustrated in Figure~\ref{fig:every_species}. 
}
\label{fig:sketch}
\end{figure}

This section presents the fundamental physical processes that govern dust dynamics, coagulation, and fragmentation in PPDs, as illustrated in Figure~\ref{fig:sketch}. We also introduce the characteristic timescales associated with these processes, which determine their relative importance in dust evolution. Lastly, we establish the notation system used throughout this study.

\subsection{Main Physical Processes}
\label{sec:St}
In PPDs, radial pressure gradients induce sub- and super-Keplerian gas rotation, causing headwind and tailwind effects that drive radial dust drift. A crucial quantity for characterizing this effect is the dimensionless stopping time, or Stokes number, ${\rm St}$. In the Epstein regime, the Stokes number is defined as:
\begin{equation}
{\rm St} = \frac{\pi a \rho_{\rm s}}{2 \Sigma_{\rm g}},
\end{equation}
where \(a\) is the particle's equivalent radius, \(\Sigma_{\rm g}\) is the gas surface density, and \(\rho_{\rm s}\) is the bulk density of the dust particle \citep{Weidenschilling_1977,Birnstiel_2010}. The dust size corresponding to \({\rm St} = 1\) is given by:
\begin{equation}
    \label{eq:a_st}
    a_{\rm St=1}=\frac{2\Sigma_{\rm g}}{\pi \rho_{\rm s}}.
\end{equation}

The dust radial drift velocity due to the pressure gradient is given by \citep{Weidenschilling_1977, Nakagawa_1986}:
\begin{equation}
    \label{eq:v_drift}
    v_{\rm drift} = \frac{{\rm St}}{1+{\rm St}^2} \frac{c_{\rm s}^2}{v_{\rm K}} \frac{\mathrm{d}\ln P}{\mathrm{d}\ln r},
\end{equation}
where \(v_{\rm K}\), \(c_{\rm s}\), and \(P\) represent the Keplerian velocity, sound speed, and gas pressure, respectively. Particles drift towards regions of higher gas pressure, leading to concentration at local pressure maxima. Generally, larger grains exhibit faster radial drift speeds when \({\rm St} \lesssim 1\).

A further important effect is the diffusion of dust grains due to turbulent gas motion. The radial dust particle diffusion coefficient, derived from the Langevin equation for particle motion in a homogeneous isotropic Kolmogorov turbulence spectrum, is \citep{Youdin2007}:
\begin{equation}\label{eq:Dp}
    D_{\rm p} = D_{\rm g} \frac{1+4{\rm St}^2}{(1+{\rm St}^2)^2}\sim \frac{D_{\rm g}}{1+{\rm St}^2},
\end{equation}
where \(D_{\rm g} = \nu / {\rm Sc} = \alpha c_{\rm s}^2 / \Omega_{\rm K} {\rm Sc}\) represents the gas diffusion coefficient, \(\Omega_{\rm K}\) is the local Keplerian angular velocity, and \(\alpha\) is the Shakura-Sunyaev turbulence parameter \citep{ShakuraSunyaev}. Here, \(D_{\rm g} = \nu\) when the Schmidt number \({\rm Sc} = 1\). Dust diffusion is more efficient for smaller particles, characterized by smaller \({\rm St}\), which leads to the spreading of dust grains concentrated at the pressure maximum.

Collisions between dust grains lead to size changes. High-energy collisions result in fragmentation \citep{Ormel_2009, Guttler_2010}, occurring when the relative velocity of grains during collisions exceeds a critical fragmentation velocity \(v_{\rm frag}\). In typical PPD conditions, turbulence provides the dominant source of relative motion for dust grains, with larger grains experiencing higher collision velocities when \({\rm St} \lesssim 1\).

The maximum Stokes number for turbulence-driven fragmentation is given by \citep{Birnstiel_2012}:
\begin{equation}\label{eq:st_frag}
{\rm St}_{\rm frag} = \frac{1}{3}\frac{v_{\rm frag}^2}{\alpha c_{\rm s}^2},
\end{equation}
which corresponds to a critical particle size \(a_{\rm frag}\) in the Epstein regime: \footnote{Note that we have omitted the correction factor \(f_{\rm frag} = 0.37\) for the fragmentation limit as suggested by \citet{Birnstiel_2012}.}
\begin{equation} \label{eq:a_frag}
a_{\rm frag} = \frac{2\Sigma_{\rm g}}{3\pi\rho_{\rm s}}\frac{v_{\rm frag}^2}{\alpha c_{\rm s}^2}.
\end{equation}

\subsection{Timescales of Growth, Diffusion, and Drift}
\label{sec:timescales}
Dust evolution in PPDs proceeds on several characteristic timescales, each governing a specific physical process.

The first is the grain growth timescale, which characterizes the rate at which dust grains grow. The typical growth timescale is given by \citep{Brauer2008a,Laune_2020,Birnstiel_2012}:
\begin{equation}
    \label{time:growth}
    \tau_{\rm growth} \sim \frac{1}{\epsilon_{\rm{tot}}\Omega_{\rm K}}, 
\end{equation}
where 
\begin{equation}
\label{def:eps}
\epsilon_{\rm{tot}}(r) \equiv \Sigma_{\rm d} / \Sigma_{\rm g}
\end{equation}
is the total dust-to-gas surface density ratio. Once the system reaches coagulation-fragmentation equilibrium, the growth timescale reflects the characteristic time for restoring perturbations.

Second, dust grains are continuously transported toward the local pressure maximum or the central star via radial drift. For dust drifting in a Gaussian pressure bump, $P(r)\propto \mathrm{exp}\left(-{0.5(r-r_0)^2}/{w_{\rm p}^2}\right)$, where $w_{\rm p}$ represents the width of the pressure bump, the drift timescale simplifies to:
\begin{equation}
    \label{time:drift}
    \tau_{\rm drift} =\left|\frac{w_{\rm p}}{\bar{v}_{\rm drift}}\right| \sim 2{\rm St}^{-1}\frac{\Omega_{\rm K}}{c_{\rm s}^2}w_{\rm p}^2.
\end{equation}
$\bar{v}_{\rm drift}$ is the averaged drift velocity within the pressure bump, and ${\rm St\ll1}$ is assumed here, which is justified in our model.

By comparing the growth and drift timescales, the drift limit determines the maximum grain size that can be replenished by growth rather than being lost due to radial drift. For dust grains within pressure bumps, this limit is \footnote{For disk particles drifting toward the central star, the corresponding drift limit is given by \citep{Birnstiel_2012}:\(a_{\rm drift,*} = \frac{2\Sigma_{\rm d}}{\pi\rho_{\rm s}}\frac{v_{\rm K}^2}{c_{\rm s}^2}\left|\frac{d\ln P}{d\ln r}\right|^{-1}.\)}:
\begin{equation}
    {\rm St}_{\rm drift} = 2\epsilon\frac{\Omega_{\rm K}^2}{c_{\rm s}^2}w_{\rm p}^2.
\end{equation}
Expressed in terms of grain size, it follows as follows:
\begin{equation}\label{eq:a_drift}
    a_{\rm drift} = \frac{4\Sigma_{\rm d}w_{\rm p}^{2}}{\pi\rho_{\rm s}}\frac{\Omega_{\rm K}^2}{c_{\rm s}^2}.
\end{equation}

The last timescale we consider is the dust diffusion timescale. 
\begin{equation}
    \tau_{\rm diff} \sim \frac{w_{\sigma_{\rm d}}^2}{D_{\rm p}},
\end{equation}
For well-coupled dust particles (i.e., ${\rm St\ll1}$), \(w_{\sigma_{\rm d}} \to w_{\rm p}\) and \(D_{\rm p}\to D_{\rm g}\), this timescale simplifies to: 
\begin{equation}
    \label{time:diff}
    \tau_{\rm diff} \sim \frac{w_{\rm p}^2}{D_{\rm g}}.
\end{equation}

\subsection{Width Notations}
\label{sec:notation}
To provide a precise quantitative description of various quantities within a pressure bump and to avoid potential confusion, we introduce a standardized notation system before proceeding with the discussion.

Within the 1D dust ring model, the radial profiles of physical quantities in the central region of a pressure bump are approximated as Gaussian functions. The standard deviation of these Gaussian profiles is defined as the corresponding width. The subscript of each width notation corresponds to the respective quantity. 

Table~\ref{table:width_summarise} provides a summary of the various ring width definitions used in this study. To characterize the dust distribution in both the radial and size domains, by distributing the total mass into logarithmic size bins, the dust size distribution is defined as:
\begin{equation}
\label{eq:def_sigma}
\sigma_{\rm d}(a,r) \equiv \frac{\partial\Sigma_{\rm d} (r)}{\partial\ln{a}}.
\end{equation}
where \(\Sigma_{\rm d}(r)\) represents the vertically integrated total dust surface density. The subscript ``\(\sigma_{\rm d},a\)'' of \(w_{ \sigma_{\rm d},a}\) is used to specify the width of the radial profile \(\sigma_{\rm d}(a,r)|_{a=a}\) for a given dust population. The same principle also applies to single species dust-to-gas ratio \(\epsilon(a,r)\).
\begin{table}[ht]
    \caption{Definitions and relationships between ring widths in radial profiles of different quantities}
    \label{table:width_summarise}
    \centering
    \renewcommand{\arraystretch}{1.4}
    \begin{tabular}{@{}lccc@{}}
    \hline\hline
    Symbol & Corresponding Quantity & Ref. \\
    \hline
    $w\mathrm{_{i}}$ &  Dust emission intensity $I(r)$  & Eq.~(\ref{eq:I_ring}) \\
     $w_{ \sigma_{\rm d},a}$ &  \parbox{4cm}{\centering Single-species \\dust surface density $\sigma \mathrm{_d}(a,r)$} & \parbox{2cm}{\centering Eq.~(\ref{eq:representation}) }\\
    $w\mathrm{_{p}}$ & Gas pressure $P(r)$ & Eq.~(\ref{eq:P_bump}) \\
    $w\mathrm{_{\epsilon}}$ &  Total dust-to-gas ratio  $\epsilon_{\rm{tot}}(r)$ & Eq.~(\ref{eq:epsilon}) \\
    $w_{\epsilon,a}$ & \parbox{4cm}{\centering Single-species\\ dust-to-gas ratio $\epsilon(a,r)$}  & \parbox{2cm}{\centering Eq.~(\ref{eq:d2gepsilon})} \\
    \hline\hline
    \end{tabular}
\end{table}
The quantity $w\mathrm{_p}$ is used as an approximation for the width of the gas surface density profile, \(\Sigma_{\rm{g}}\), as adopted in this study and in \cite{Dullemond_2018}. This approximation holds if the temperature variation across the bump is negligible.

By balancing dust drift and diffusion, while neglecting dust growth and fragmentation, the intrinsic width of dust rings in surface density for a \textit{single} dust species, \( w_{\sigma_{\rm d},a} \), is given by \citep{Dullemond_2018, Morbidelli_2020}.
\begin{equation}
\label{eq:Dullemond}
   w_{\sigma_{\rm d},a} = w_{\rm p}(1 + \psi^{-2})^{-1/2},
\end{equation}
where \( \psi = \sqrt{\alpha / (\mathrm{St_0}(a) \cdot \rm{Sc})} \), \( \mathrm{St_0} \) (which scales with the dust size $a$) is the Stokes number at the bump center, and \( \mathrm{Sc} \) is the Schmidt number, typically assumed to be \( \mathrm{Sc} = 1 \).
For large particles where \( \psi \ll 1 \), it follows that \( w_{\sigma_{\rm d},a} \simeq w_{\rm p} \psi \), suggesting dust trapping. Conversely, for small particles where \( \psi \gg 1 \), \( w_{\sigma_{\rm d},a} \simeq w_{\rm p} \), indicating that small dust grains are well-coupled with the gas. 

It is essential to distinguish between \(w_{\epsilon,a}\) and \(w_{\sigma_{\rm d},a}\), as both describe the dust distribution from different perspectives. The dust surface density \(\sigma_\mathrm{d}\) directly characterizes the spatial distribution, whereas the dust-to-gas ratio \(\epsilon\) is dynamically more relevant, governing the net dust diffusion flux. These two representations are related by \(\sigma_{\rm{d}}({a},r) = \epsilon({a},r) \cdot \Sigma_{\rm{g}}(r)\), leading to the following relation:
\begin{equation}
\label{eq:representation}
    w_{\sigma_{\rm d},a}^2 = \frac{w_{\epsilon,a}^2\cdot w_{\rm p}^2}{w_{\epsilon,a}^2 + w_{\rm p}^2}.
\end{equation}

By combining Eqs.~\eqref{eq:Dullemond} and \eqref{eq:representation}, we have the ring width in terms of the dust-to-gas ratio,
\begin{equation}
\label{eq:Dullemond_eps}
   w_{\epsilon,a} = w_{\rm p} \sqrt{\frac{\alpha}{\mathrm{St_0}(a)}}.
\end{equation}
This is the drift-diffusion equilibrium solution \citep{Dullemond_2018}.

In the limit of well-coupled dust with ${\rm St_{0}}\ll1$, \(w_{\epsilon,a} \to \infty\). 
The corresponding ring width for the dust surface density \(w_{\sigma_{\rm d},a}\) asymptotically approaches \(w_{\rm p}\),

\section{Numerical Simulation Method} 
\label{sec:sim_setup}

We simulate the dynamical evolution of a typical dust ring in the PPD using the open-source \texttt{Python} code \texttt{DustPy} \citep{Stammler_2022}. The simulation begins with a gas pressure bump embedded in the disk, evolving multiple dust species under coagulation, fragmentation, drift and diffusion processes.

\subsection{Disk Setup}

\begin{figure}
\includegraphics[width=\columnwidth]{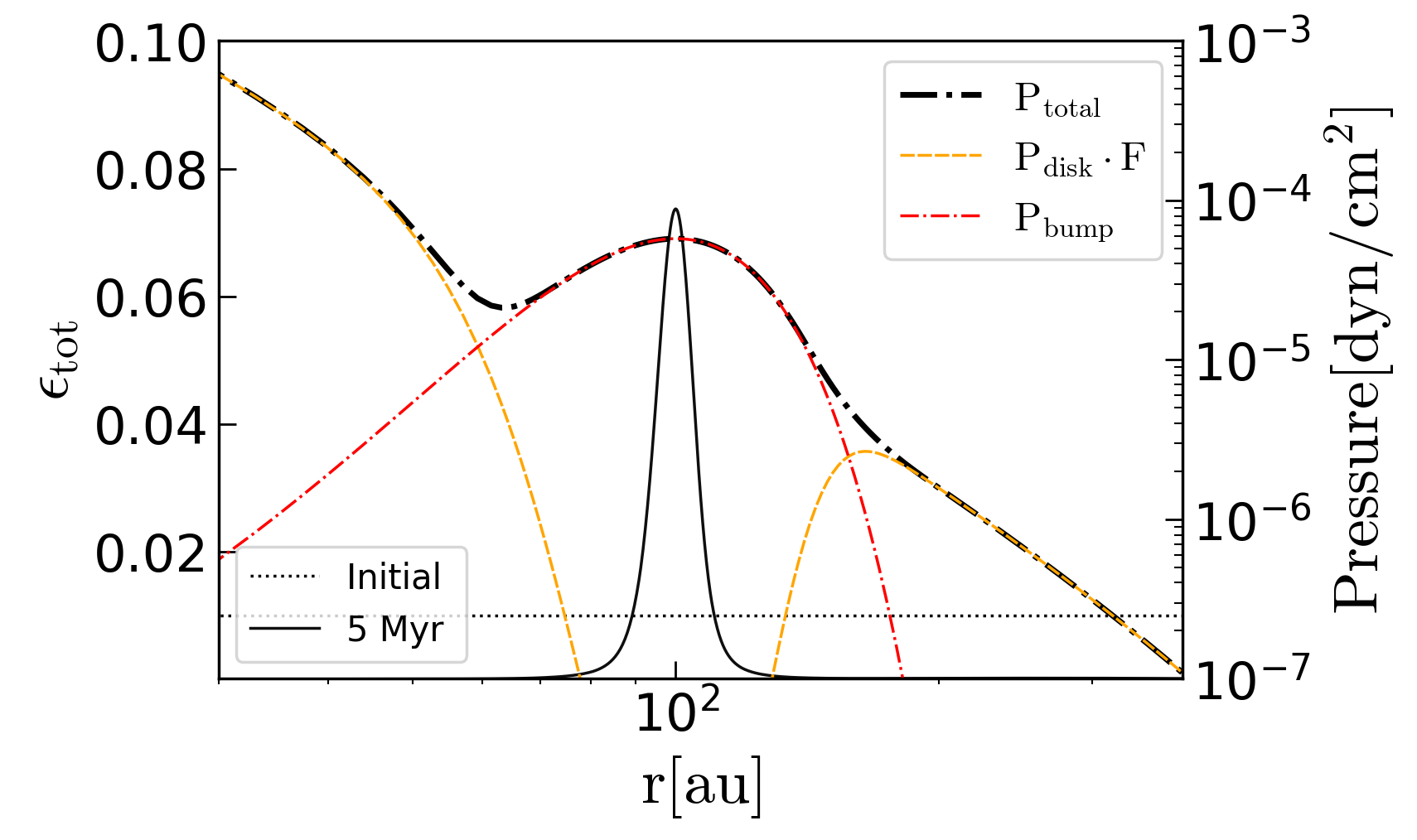}
\caption{Radial profiles of the total gas pressure \( P_{\rm total} \) (right y-axis) and the dust-to-gas ratio \( \epsilon \) (left y-axis). The colored lines (right y-axis) represent the two terms on the right-hand side of Eq.~\eqref{eq:p_total}, respectively. The pressure profile remains fixed throughout the simulations. The dotted line represents the initial dust-to-gas ratio, while the solid line corresponds to the profile at 5.0 Myr, which is close to the steady state. The steady-state \( \epsilon \) is well approximated by a Gaussian bump, allowing us to measure its width \( w_{\rm e} \).}
\label{fig:eps}
\end{figure}

The total pressure profile is prescribed as a pressure bump superimposed on a minimized pressure background, expressed as:
\begin{equation}
\label{eq:p_total}
P_{\rm total}(r) = P_{\rm bump}(r) + P_{\rm disk}(r) F(r),
\end{equation}
where \( P_{\rm bump}(r) \) and \( P_{\rm disk}(r) \) represent the pressure profiles of the Gaussian bump and the smooth background, respectively. The factor \( F(r) \), detailed below, further modifies the background pressure. These components are illustrated in Figure~\ref{fig:eps}.

The pressure bump is initialized with a Gaussian profile:
\begin{equation}
    \label{eq:P_bump}
    P\mathrm{_{bump}}(r)=P\mathrm{_{bump,0}}\cdot \mathrm{exp}\left(-\frac{(r-r_0)^2}{2w_{\rm p}^2}\right),
\end{equation}
where \( w_{\rm p} \) denotes the width of the pressure bump. The peak mid-plane pressure is given by \( P\mathrm{_{bump,0}}= \frac{c_{\rm s,0}^2\Sigma_{\rm g,0}}{\sqrt{2\pi}H_{\rm p,0}} \), where \( H_{\rm p} \) is the pressure scale height, and the subscript ``0'' indicates values measured at the pressure maximum \( r_{0} \).

The smooth background disk pressure is defined as:
\begin{equation}
P_{\rm disk}(r)=\frac{c_{\rm s}^{2}(r)\Sigma_{\rm disk}(r)}{\sqrt{2\pi}H_{\rm p}(r)}, 
\end{equation}
where the surface density profile \( \Sigma_{\rm disk}(r) \) follows a cutoff power-law profile \citep{Lynden-Bell1974}:
\begin{equation}
\Sigma_{\rm disk}(r)=\frac{\Sigma_{\rm g,0}}{A_{\rm p}} \left(\frac{r}{r_{0}}\right)^{-1}\exp\left(-\frac{r}{r_{\rm c}}\right),
\end{equation}
where \( A_{\rm p} \) quantifies the contrast between the background disk and the pressure bump, satisfying \( P_{\rm bump}(r_0)  = A_{\rm p} P_{\rm disk}(r_0) \). The parameter \( r_{\rm c} \) represents the critical cutoff radius of the disk. The temperature follows \(T=T_0(r/r_0)^{-1/2}\), resulting in a radial sound speed profile \( c_{\rm s} \propto r^{-1/4} \) \citep{Chiang_Goldreich_1997}. 

To facilitate analytical treatment, the function
\begin{equation}
F(r) \equiv \exp\left(-f \exp\left(-\frac{(r - r_0)^2}{2w_{\rm p}^2}\right)\right)
\end{equation}
is introduced to preserve the Gaussian form of the \( P_{\rm bump}(r) \) and ensure a well-defined pressure bump width \( w_{\rm p} \). We set \( f = 9.3 \) to maintain the pressure bump almost unchanged.

For consistency, we select fiducial model parameters appropriate for the ring at 100 au in the HD 163296 system, as summarized in Table~\ref{table:setup}. The total gas pressure profile \( P_{\rm total} \) is shown as the dotted-dashed line in Figure~\ref{fig:eps}, demonstrating that the Gaussian bump remains largely intact without significant alteration from the background pressure.

Our analysis specifically targets moderate-strength pressure bumps exhibiting two critical properties: (1) a local pressure maximum (${\rm d}P/{\rm d}r=0$) in the total pressure profile, and (2) being stable against Rossby-wave instability (RWI).
First, this scenario aligns with observational evidence for pressure maxima in dust-trapping rings \citep[e.g.,][]{rosotti_2020}. Second, this is theoretically supported by our model's robust stability against RWI across an extensive parameter space (detailed in Appendix~\ref{app:rwi}).
In contrast, pressure bumps lacking a local maximum produce fundamentally different dust dynamics: radial drift dominates over growth processes. This yields size-dependent ring centers and analytically intractable equilibrium solutions for dust ring width. This scenario is beyond this study's scope and needs future detailed studies.

For simplicity, we neglect viscous accretion in the disk and assume the gas pressure profile remains constant over time. In this context, the back-reaction of dust on the gas is not considered in this study.

\begin{table}[ht]
\caption{Fiducial simulation parameters, corresponding to the ring at 100 au in HD 163296.}
\label{table:setup}
    \centering
    \begin{tabular}{@{}lccc@{}}
    \hline\hline
    Symbol & Description & Value & Unit \\
    \hline
    $M_*$ & Stellar mass & 1.9 & $M_{\odot}$ \\
    $r\mathrm{_c}$ & Critical cutoff radius & 200 & au \\
    $r_{0}$ & Pressure maximum position& 100.0 & au \\
    $A\mathrm{_{p}}$ & Pressure bump contrast & 2 & --- \\
    $w\mathrm{_{p}}$ & Pressure bump width & 23 & au \\
    $\epsilon\mathrm{_{ini}}$ & Initial dust-to-gas ratio & 0.01 & --- \\
    $\rho_{\rm s}$ & Particle bulk density & 1.67 & g cm$^{-3}$ \\
    $a\mathrm{_0}$ & Minimum dust size & $5\times10^{-5}$ & cm \\
    $v_{\rm frag}$ & Fragmentation velocity & 500 & cm s$^{-1}$ \\
    $\xi$ & Power-law index of fragments & $11/6$& --- \\
    $\alpha$ & Viscosity parameter & 0.001 & --- \\
    $T_{0}$ & Temperature at $r_{0}$ & 12.3 & $K$ \\
    $\Sigma\mathrm{_{g,0}}$ &  Gas surface density at $r_{0}$ & 25 & g cm$^{-2}$ \\
    $H_{\rm p,0}$ &  Gas scale height at $r_{0}$ & 5.1 & $\rm{au}$ \\
    \hline\hline
    \end{tabular}
\end{table}
\subsection{Dust Coagulation and Fragmentation}

The dust component is initialized as micron-sized grains, uniformly distributed in the range of $0.5$ to $1 \, \mathrm{\mu m}$, with a global dust-to-gas ratio set at $\epsilon\mathrm{_{ini}}\equiv \Sigma\mathrm{_{d,ini}}/\Sigma_{\rm g} = 0.01$ (shown as the dashed line in Figure~\ref{fig:eps}). 

Dust grains moving at different velocities undergo collisions, leading to various outcomes. The relative velocities between dust grains arise from turbulence, Brownian motion, vertical settling, and radial/azimuthal drift, with turbulence typically being the dominant contributor. The collision outcomes considered in our simulations include coagulation, fragmentation, and erosion (We adopt the default coagulation model implemented in \texttt{DustPy}, as described in \cite{Stammler_2022} for further details). Fragmentation becomes increasingly likely as the collision velocity exceeds a critical threshold, i.e., the fragmentation velocity \( v_{\rm frag} \); otherwise, grains are more likely to adhere, leading to coagulation.

Upon fragmentation, the resulting fragment mass distribution, \( n_{\rm frag}(m) \propto m^{-\xi} \), follows a power law with an index of \( \xi = 11/6 \) \citep{Brauer2008a}, corresponding to an MRN-like size distribution \citep{Mathis1977}.

The simulation employs a logarithmic mass grid with a mass resolution of seven bins per decade, a configuration extensively tested in previous works for collision modeling \citep{Ohtsuki1990, Draz_2014,Stammler_2022}.

\subsection{Disk Parameters}

The other parameters for the fiducial simulation are based on the 100-au dust ring in HD 163296, as summarized in Table~\ref{table:setup}. Given a stellar mass of \( M_* = 1.9 M_{\odot} \) \citep{Fairlamb_2015} and a pressure bump width of \( w_{\mathrm{p}} = 23 \)~au \citep{rosotti_2020}, the temperature at the pressure maximum is set to \( T_0 = 12.3~\mathrm{K} \) \citep{Guidi_2022}. The disk cutoff radius is chosen as \( r_{\rm c} = 200 \)~au \citep{Stammler_2019}. Reported values for the gas surface density vary considerably across different studies (i.e., \(\sim 5\ \rm g/cm^2\) by \cite{Zagaria_2023} or \(\sim20\ \rm g/cm^2\) by \cite{rab_2020}), however it does not change our conclusions and we just set an arbitrary value for $\Sigma_{\rm g,0}$ in simulations.

The radial domain of our simulations extends from 10~au to 400~au with a grid resolution of \( \sim0.26\ \mathrm{au} \) (corresponding to \( 0.05 H_{\rm p} \)) at the bump center. A constant gradient is applied to the inner boundary and a floor value is used for the outer boundary condition.
The large radial domain of the disk with an exponential disk profile ensures that the dust dynamics around the ring is not influenced by the boundary condition.


\section{Simulation Results} \label{sec:sim_result}

Here we show the simulation results for a typical dust-trapping ring, by characterizing their size distribution and spatial distribution within the ring.

\subsection{Dust Dynamics within the Ring}
\begin{figure}
\includegraphics[width=\columnwidth]{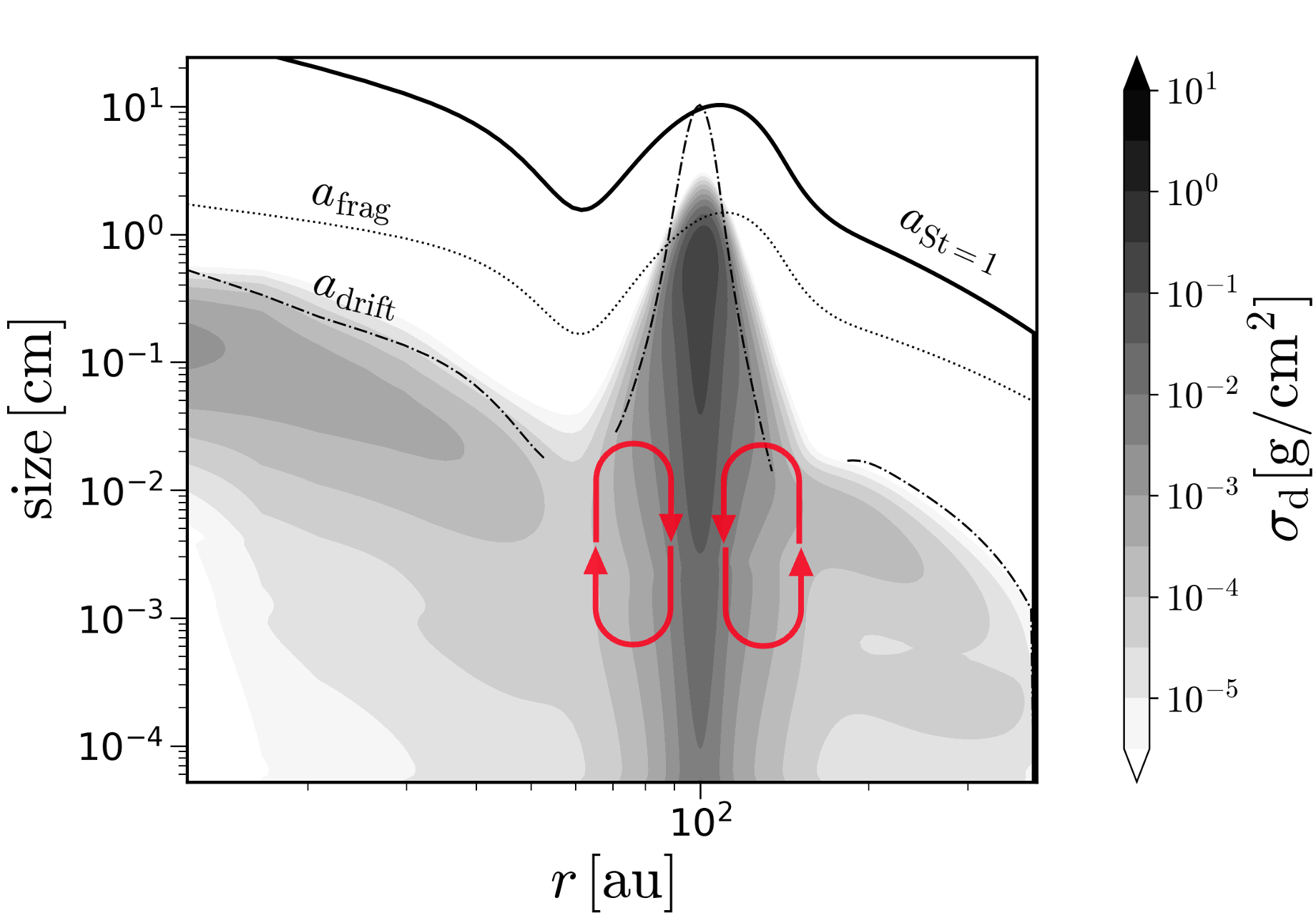}
\caption{The 2D distribution of dust surface density $\sigma\rm_d$ as a function of the orbit radius and dust size at the steady state in the fiducial simulation, 5 Myr. The critical sizes $a_{\rm drift}$, $a_\mathrm{frag}$, and $a_\mathrm{St=1}$, correspond to the drift limit, fragmentation limit, and the dust size with a unity Stokes number, respectively.  The maximum size is set by $a_\mathrm{frag}$ at the pressure bump center and by $a_{\rm drift}$ at the wings. The red arrow loop represents the ``dust cycle'', as introduced in Figure \ref{fig:sketch}, to cycle dust in this coagulation-fragmentation equilibrium.
}
\label{fig:fiducial}
\end{figure}
Figure \ref{fig:eps}  presents the steady-state (5 Myr) total dust-to-gas ratio \(\epsilon_{\mathrm{tot}}\) as a function of radius in the disk (the solid line). The total dust-to-gas ratio profile can be described by a Gaussian function 
\begin{equation}
\label{eq:epsilon}
    \epsilon_{\rm tot}(r) = \epsilon_0 \exp{\left(-\frac{(r-r_0)^2}{2w_{\epsilon}^2}\right)},
\end{equation}
from which we define the intrinsic dust ring width \(w_{\epsilon}\) (to distinguish from the width of observed intensity rings $w_{\rm i}$). We can clearly identify significant dust trapping with \(w_{\epsilon} < w_{\rm p}\) in the fiducial run, which is also evident from the red dashed line in Figure~\ref{fig:every_species}. 

Figure \ref{fig:fiducial} shows the steady-state (5 Myr) dust size distribution $\sigma_{\rm d}$. The bump region does not evolve significantly after about \(\mathrm{1\ Myr}\). As is shown, dust particles encounter two barriers during their dynamical evolution, one is the drift barrier $a_{\rm drift}$, another is the fragmentation barrier $a_\mathrm{frag}$. It is evident that the maximum size at the center of the pressure bump is primarily restricted by $a_\mathrm{frag}$, whereas the edges of the bump are constrained by $a_{\rm drift}$, as also reported by \cite{Li2019,Jiang_2024}.

\subsection{A Similar Radial Distribution for Different Dust Species within a Ring}
\begin{figure}
\includegraphics[width=\columnwidth]{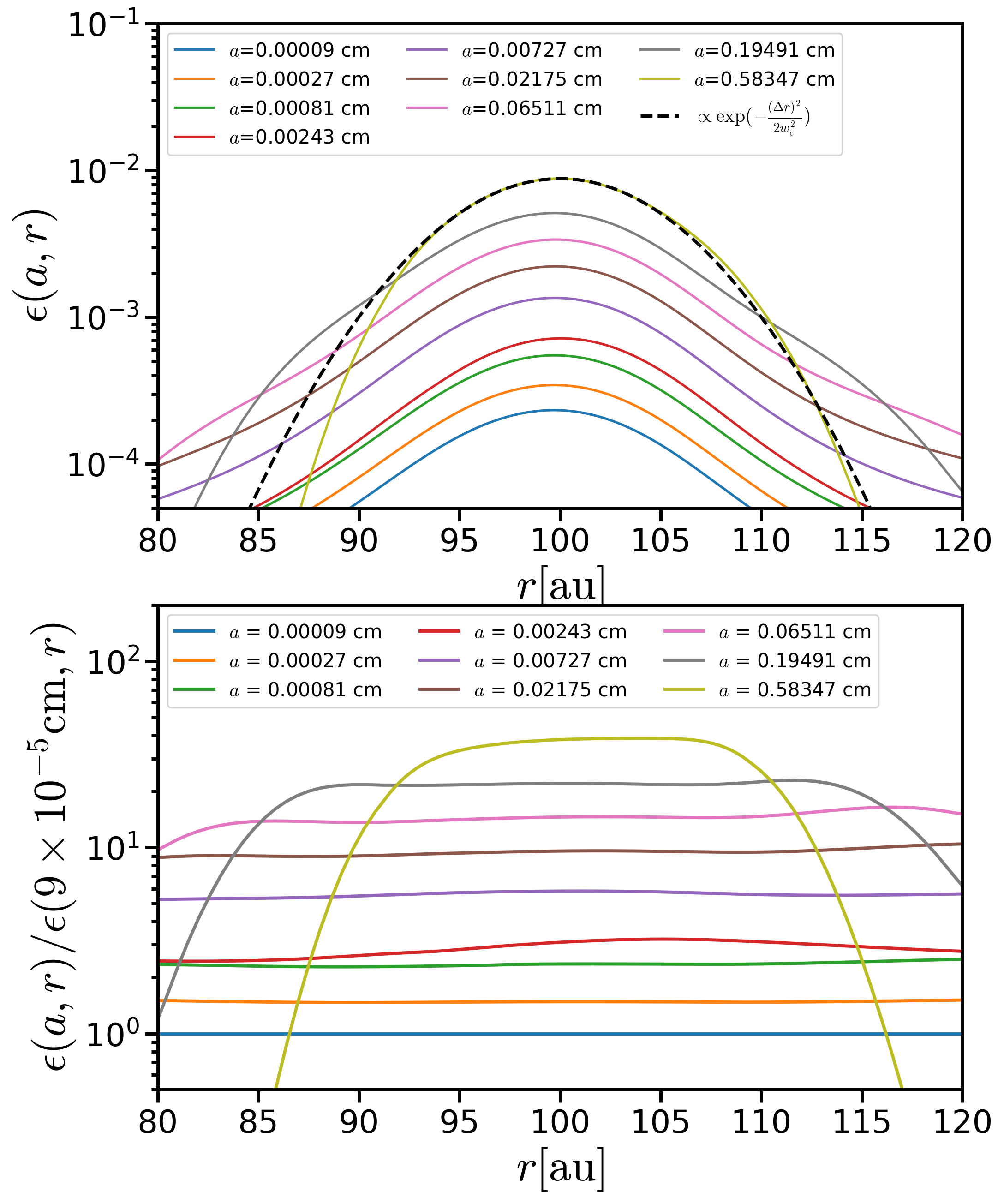}
\caption{
The upper panel shows the dust-to-gas ratio profile \(\epsilon(a,r)\) at 5 \(\rm Myr\) for a variety of gain sizes up to the fragmentation limit in the fiducial simulation.
The lower panel depicts the ratio of \(\epsilon(a,r)/\epsilon(0.00009\ \mathrm{cm},r)\) for different species. 
As is shown here, Within the dust ring region \((r_0-w_{\epsilon},r_0+w_{\epsilon})\) $[\sim 95-105\ {\rm au}]$, \(\epsilon(a,r)\) can be well approximated by a Gaussian profile. This functional form remains consistent across all particle sizes \(a<a|_{\rm St=0.37St_{\rm frag}}\), as revealed by the nearly constant \(\epsilon(a,r)/\epsilon(0.00009\ \mathrm{cm},r)\) in this region.
\label{fig:eps_profile}
}
\label{fig:eps_ratios}
\end{figure}
\begin{figure}
\includegraphics[width=\columnwidth]{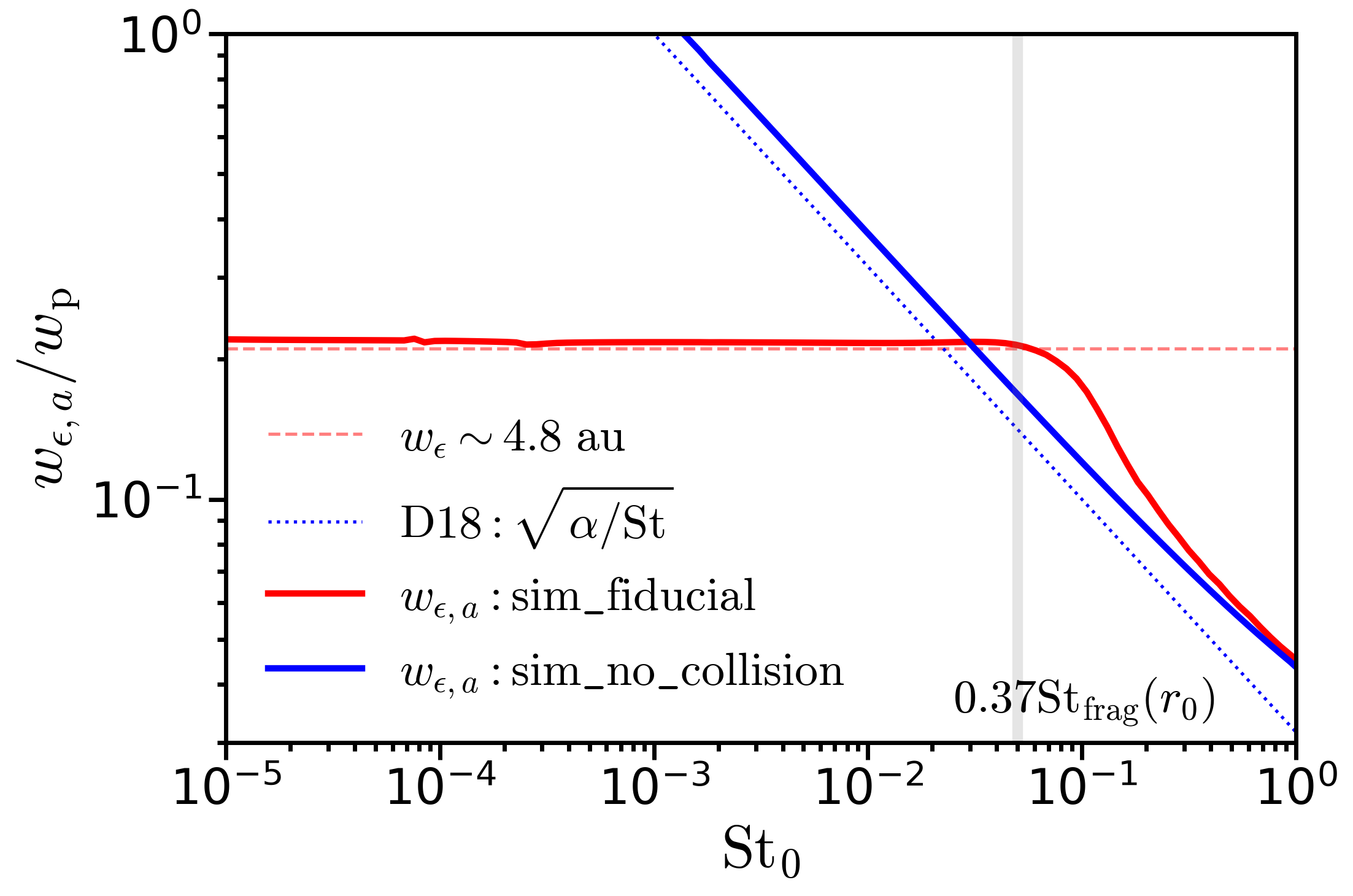}
\caption{The width of each individual dust species scaled by the bump width as a function of the dust Stokes number \(\mathrm{St}_0 \equiv \pi a \rho_{\rm s} / 2 \Sigma_{\rm g, 0}\). When dust coagulation and fragmentation are included, all dust species with ${\rm St_{0}\lesssim 0.37St_{\rm frag}}$  converge to a similar distribution, resulting in the same ring width for different species, as indicated by the red solid line. The red dashed line shows the width \(w_{\rm e}\) for the total dust-to-gas ratio \(\epsilon\) profile. In the drift-diffusion-only case (blue lines), where the dotted line represents predictions from Eq. (\ref{eq:Dullemond}) and the solid line shows the simulation results, the ring width depends strongly on dust size. The vertical gray line indicates the fragmentation barrier.}
\label{fig:every_species}
\end{figure}

Now, we examine the spatial distribution of the dust ring. Figure \ref{fig:eps_ratios} illustrates the single species dust-to-gas ratio \(\epsilon(a,r)\) for various dust sizes \(a\):
\begin{equation}
\label{eq:def_eps}
\epsilon(a,r) \equiv \frac{\sigma_{\rm d}(a,r)}{\Sigma_{\rm g}(r)}.
\end{equation}

Based on the fitting of the radial distribution of the dust-to-gas ratio, its radial profile for each dust species can be well described by:
\begin{equation}
\label{eq:d2gepsilon}
\epsilon(a,r)  \propto \exp{\left(-\frac{(r - r_0)^2}{2 w_{\epsilon,a}^2}\right)}.
\end{equation}
Figure \ref{fig:every_species} shows the measured widths \(w_{\epsilon,a}\) for each dust species as a function of \(\mathrm{St}_0\), where \(\mathrm{St}_0 \equiv \pi a \rho_{\rm s}/2 \Sigma_{\rm g,0}\) is the Stokes number at the pressure maximum, which can be interpreted as the grain size. 
\(w_{\epsilon,a}\) is derived by fitting the radial profile of \(\epsilon(a,r)\) for each species with a Gaussian profile as in Eq.~\eqref{eq:d2gepsilon}.

For dust rings regulated by coagulation and fragmentation equilibrium, the ring widths \(w_{\epsilon,a}\) are nearly identical across dust species with $\mathrm {St_0\lesssim 0.37St_{frag}}(r_0)$ in our simulations, as indicated by the red solid line. 
This is further confirmed by direct comparison of radial dust-to-gas ratio profiles $\epsilon(a,r)$ across particle sizes as shown in Figure~\ref{fig:eps_profile}, which exhibit consistent shapes differing only in normalization -- matching our expectation.

This contrasts with the drift-diffusion scenario, as illustrated by the blue lines in Figure \ref{fig:every_species} and discussed by \citet{Dullemond_2018}. If dust collisions are neglected (i.e., only drift and diffusion are considered), the width \( w_{\epsilon,a} \) becomes size-dependent. Larger, drift-dominated grains are more tightly concentrated, whereas smaller, diffusion-dominated particles are less effectively trapped. Consequently, for well-coupled dust, \( w_{\sigma_{\rm d},a} / w_{\rm p} \simeq 1 \), which corresponds to \( w_{\epsilon,a} / w_{\rm p} \propto \sqrt{1/{\rm St}} > 1 \), as explained in Section~\ref{sec:notation}.

However, grains near the fragmentation limit, \(\mathrm{St}_{\rm frag}\), exhibit a narrower width than \(w_{\epsilon}\), as shown in Figure~\ref{fig:every_species}. This narrowing arises primarily from two factors. 
First, for even larger pebbles with shorter drift timescales, drift-diffusion becomes the dominant process, and their distribution follows the drift-diffusion equilibrium solution (blue lines in Figure~\ref{fig:every_species}). This explains why the red line converges with the blue line for very large pebbles. 
Second, the fragmentation limit varies with orbital radius, restricting grains near this threshold to a confined region. Specifically, in the central part of the dust ring, the fragmentation limit $a_{\rm frag}$ decreases outward from the ring center (as indicated by the dotted line in Figure~\ref{fig:fiducial}), allowing \(\sim 1\,\mathrm{cm}\) pebbles to exist only near the center. Consequently, these pebbles are more spatially constrained compared to smaller grains. 
Together, the two mechanisms lead to a narrower radial distribution for large pebbles, even though we expect that the first effect could be more important. 

Although the effect above has a negligible impact on dust dynamics -- since the fraction of dust deviating from the common radial distribution is small -- it could significantly affect thermal radiation, providing an observable signature that can be used to test the theory, as discussed in the following paper.

\subsection{A Similar Dust Size Distribution for Different Radii within a Ring}
\begin{figure}
\includegraphics[width=\columnwidth]{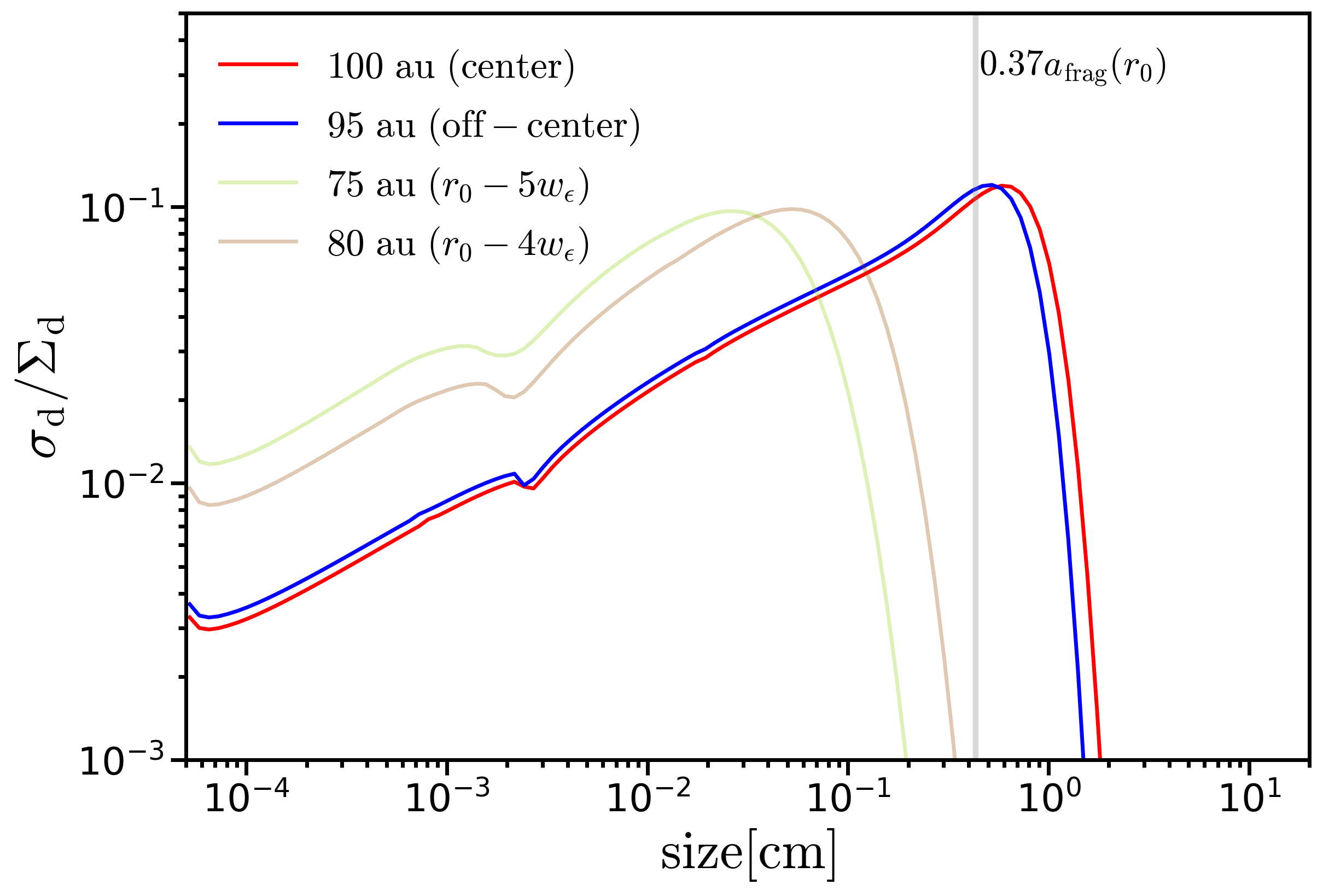}
\caption{
The normalized dust size distribution \(\sigma_\mathrm{d}/\Sigma_\mathrm{d}\) as a function of dust size. The red solid line represents the distribution at the bump center (100 au) at steady state, while the blue solid line corresponds to an off-center location at \(\sim 95\) au. The vertical gray line marks the fragmentation limit at the bump center. For comparison, distributions at the bump wing (\(\sim 75\) au and \(\sim 80\) au) are shown as semi-transparent lines. The near-independence of \(\sigma_\mathrm{d}/\Sigma_\mathrm{d}\) with radius in the dust ring region suggests a coagulation-fragmentation equilibrium.
}
\label{fig:dust_distribution}
\end{figure}

Next, we describe the dust size distribution within the central region of the bump. Figure \ref{fig:dust_distribution} presents the normalized dust size distribution \(\sigma_{\rm d}(r) / \Sigma_{\rm d}\) as a function of dust size at different radii represented by the solid (the dust ring at 100 and 95 au) and semi-transparent (the bump wing at 80 and 75 au) lines, respectively. The size distributions in the dust ring ($95-105\ {\rm au}$) closely matches with each other, indicating a similar dust size distribution in the dust ring region, which is notably different from the size distribution in the wings.

The reason for this similar dust size distribution is as follows. The normalized size distribution \(\sigma_{\rm d} / \Sigma_{\rm d}\) is primarily characterized by its slope and maximum size. In the case that coagulation-fragmentation is efficient enough that the slope for the dust size distribution is governed by the spatially-independent fragmentation slope \(\xi\) \citep{Stammler_2022}, while the maximum size, determined by turbulence-driven fragmentation, is associated with a spatially invariant Stokes number \(\mathrm{St}_{\rm frag}\) (see Eq.~\ref{eq:st_frag}). Together, these factors suggest that the distribution \(\sigma_{\rm d} / \Sigma_{\rm d}\) remains spatially invariant if the dust dynamics are primarily governed by coagulation-fragmentation equilibrium. In contrast, the inclusion of dust drift and diffusion, both of which depend on spatial position, would result in significant variations in \(\sigma_{\rm d} / \Sigma_{\rm d}\) within the bump.
The similar dust size distribution thus indicates a coagulation-fragmentation equilibrium in the dust ring, and suggests that radius-dependent drift and diffusion effects can be neglected in terms of the dust size distribution.

The vertical gray line in Figure \ref{fig:dust_distribution} represents the predicted fragmentation barrier at the bump center \(a_{\rm frag}\), corrected by a factor of \(0.37\). The simulation results show that dust accumulates around a slightly smaller size than \(a_{\rm frag}\). This discrepancy arises due to the implementation of the Maxwell-Boltzmann distribution for dust-relative velocities \citep[see Eq. 58 in][]{Stammler_2022}, which is not considered in the analytical analysis. The high-velocity tail of the Maxwell-Boltzmann distribution results in fragmentation occurring at a smaller size.

The size-independent ring width \(w_{\epsilon,a}\) aligns with the spatially-insensitive size distribution \(\sigma_{\rm d}/\Sigma_{\rm d}\) across different radii, both of which indicate a coagulation-fragmentation equilibrium in the dust ring, where the majority of the dust mass resides.

In summary, different dust species within the ring exhibit a similar size and spatial distribution. This suggests that the size distribution at various radii within the ring can be described by the same function, i.e., 
\[
\left(\frac{\sigma_{\rm _d}}{\Sigma_{\rm d}}\right)\bigg|_r = \left(\frac{\sigma_{\rm _d}}{\Sigma_{\rm d}}\right)\bigg|_{r_0}\gets \substack{\text{coagulation/fragmentation } \\ \text{equilibrium}}.
\]
Furthermore, the radial width of the ring for different dust species within the fragmentation limit is represented by a common constant, i.e., 
\[
w_{\epsilon, a<0.37a_{\rm frag}} = w_{\epsilon}.
\]

\section{The Dust Ring Model} \label{sec:analytical_fits}

In this section, we present the analytical reconstruction of the dust ring based on a comprehensive parameter survey and calibration from our simulation results.

\subsection{Understanding the Dust Ring with Coagulation: An Analytical Framework}

We skip the discussion about dust size distribution in coagulation-fragmentation equilibrium, \(\sigma_{\rm d}\), which is well-studied by \cite{Birnstiel_2011}. 

The width of the dust-to-gas ratio profile for a dust ring, $w_{\epsilon}$ is given by:
\begin{equation}
    \label{eq:wewg}
    w_{\epsilon} = 1.2 \, w_{\rm p} \cdot \sqrt{\frac{\bar{D}}{\bar{V}}}.
\end{equation}
The factor of 1.2 is a constant calibrated from simulation results. Here, \(\bar{V}\) represents the average drift effect and \(\bar{D}\) represents the average diffusion effect for the size distribution at the bump center \(\sigma_{\rm d}|_{r_0}\), and are computed as
\begin{equation}
\label{eq:vbar}
    \bar{V} \equiv \int_{a_0}^{\infty} \left( \sigma_{\rm d} \frac{\mathrm{St}}{1 + \mathrm{St}^2} \right)\Bigg|_{r_0} \,d\ln a,
\end{equation}
and
\begin{equation}\label{eq:dbar}
    \bar{D} \equiv \int_{a_0}^{\infty} \left( \sigma_{\rm d} \frac{\alpha}{1 + \mathrm{St}^2}\right)\Bigg|_{r_0}  \,d\ln a.
\end{equation}
A detailed derivation of $w_{\epsilon}$ is provided in Appendix \ref{append:analytical_we}. 

As an illustrative example, consider a power-law size distribution of dust grains, \(n(a) \propto a^{-q}\), with a maximum grain size \(a\mathrm{_{max}}\). If the Stokes number of the largest grain, \(\mathrm{St_{max}}\), is less than unity, \(\mathrm{St_{max}} < 1\), \(a_{\mathrm{min}} \ll a_{\mathrm{max}}\) and \(q < 4\), the following approximation holds:
\begin{equation}
    \frac{w_{\epsilon, a<0.37a_{\rm frag}}}{w_{\rm{p}}}=\frac{w\mathrm{_\epsilon}}{w_{\mathrm{p}}} = 1.2 \sqrt{\frac{5 - q}{4 - q} \frac{\alpha}{\mathrm{St_{max}}}}.
\end{equation}
The ratio \(w_{\epsilon}/w_{\rm{p}}\), along with widths for smaller grains of arbitrary size, is determined by \(\mathrm{St_{max}}\). In the fragmentation-limited regime, where \(\mathrm{St_{max}} = \mathrm{St_{frag}}\), \(w_{\epsilon, a<0.37a_{\rm frag}}\) becomes independent of the gas surface density \(\Sigma_{\rm g}\), and the dust bulk density \(\rho_{\rm s}\) (see Eq.~\ref{eq:st_frag}). This behavior contrasts with the drift-diffusion scenario, in which the ring width \(w_{\epsilon, a<0.37a_{\rm frag}}\) for a given dust species scales as \(\sqrt{1/\mathrm{St}(a, \Sigma_{\rm{g}}, \rho_{\rm{s}})}\) \citep{Dullemond_2018}.

\subsection{Verification of Eq. (\ref{eq:wewg})}
\begin{figure*}
\includegraphics[width=\textwidth]{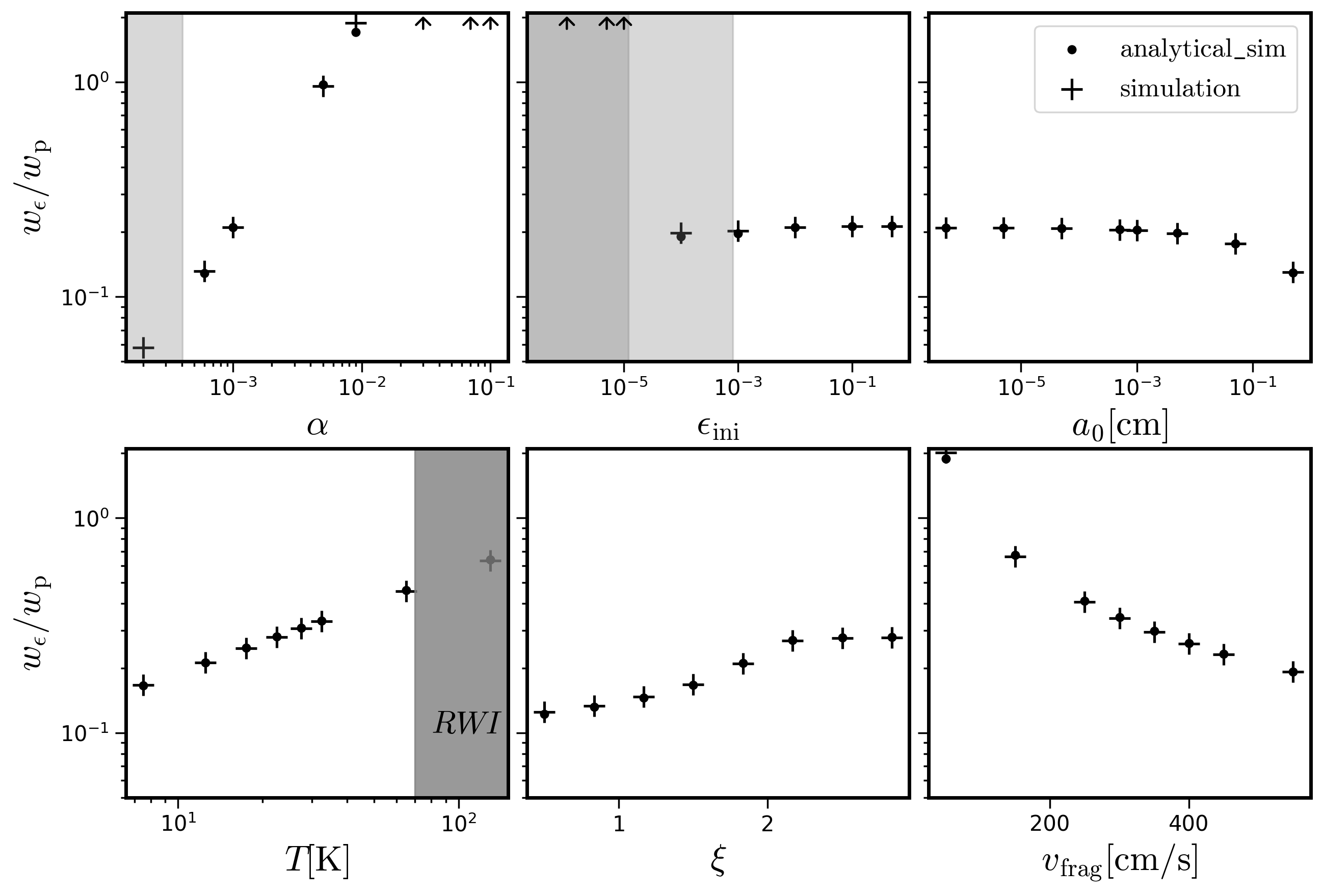}
\caption{
The dust ring width, scaled by the bump width \(w_\mathrm{\epsilon}/w_\mathrm{p}\), is shown across the parameter space, with each panel varying one model parameter from the fiducial simulation. The '+' markers represent direct measurements of the intrinsic dust ring width in simulations at steady state, while the dots indicate values calculated using Eq.~(\ref{eq:wewg}), with \(\sigma_\mathrm{d}\) derived from the corresponding simulations, demonstrating excellent agreement under typical conditions. The shaded regions highlight the parameter space where the model is no longer valid.}
\label{fig:width_para}
\end{figure*}
Eq. (\ref{eq:wewg}) is verified through comparison with simulation results from an extensive parameter survey. We performed simulations by varying one parameter at a time from the fiducial model and measuring the steady-state intrinsic dust ring width, as given by Eq.~(\ref{eq:epsilon}). The parameters explored include the disk turbulence parameter \(\alpha\), the initial dust-to-gas ratio \(\epsilon_{\rm ini}\), the initial monomer dust size \(a_{0}\), disk temperature \(T\), the power-law index of fragments \(\xi\), and the fragmentation velocity \(v_{\rm frag}\). 

We then calculated the ring width using Eq. (\ref{eq:wewg}), applying the dust size distributions, \(\sigma_{\rm d}|_{r_0}\) directly taken from the simulations.

The measured and calculated ring widths are shown in Figure \ref{fig:width_para}. The shaded regions indicate the parameter space where the analytical theory is no longer applicable. Specifically, for very small values of \(\alpha\), which correspond to a laminar disk environment, the relative velocity between dust grains never reaches \(v_{\rm frag}\), preventing fragmentation. This scenario is out the scope of the current coagulation model \citep{Birnstiel_2009}. Additionally, when the dust-to-gas ratio is extremely low, drift and diffusion effects dominate, as discussed in Appendix \ref{append:analytical_we} (Eqs. \ref{diff_criteria} and \ref{drift_criteria}), requiring alternative treatment (Appendix \ref{appdenix:low_epsilon_approxi}). 

The ring width calculated using the simulation-derived \(\sigma_{\rm d}\) (represented by the dots) matches the measured values perfectly, confirming the validity of the ring width formula described by Eq.~(\ref{eq:wewg}). 

\subsection{Reconstructing the Dust Ring}

We assume that the dust ring is in drift/diffusion/coagulation/fragmentation equilibrium. And the dust-to-gas ratio is high enough to meet the criterion Eq.~\eqref{drift_criteria}. The main steps to reconstruct the dust ring are summarized below, starting from a gas pressure bump:

\begin{description}
    \item[Step 1] Construct the normalized dust size distribution \((\sigma_{\rm d}/\Sigma_{\rm d})|_{r_0}\) at the pressure maximum based on the fragmentation limit \(a_{\rm frag}\) and the power-law indices for multi-species dust \citep[see][for details]{Birnstiel_2018,Birnstiel_2011}.
    
    \item[Step 2] Calculate \(w_{\epsilon,a}\) for each dust species. First, use the gas bump width \(w_{\rm p}\) and the dust size distribution \((\sigma_{\rm d}/\Sigma_{\rm d})|_{r_0}\) to determine the dust ring width \(w_{\epsilon}\) for the total dust-to-gas ratio \(\epsilon\), based on Eq.~(\ref{eq:wewg}). This width applies to dust species with sizes of \({a<0.37a_{\rm frag}}\). For larger grains, the drift-diffusion balance solution Eq. (\ref{eq:Dullemond_eps}), scaled by the factor \((w_{\epsilon}/w_{\rm p})(\sqrt{\mathrm{St}(0.37a_{\rm frag})/\alpha})\), provides a good approximation.
    
    \item[Step 3] Map the 2D size-spatial distribution of dust, similar to that shown in Figure \ref{fig:fiducial}, by distributing \(\sigma_{\rm d}(a,r) = \Sigma_{\rm g}(r)\cdot\epsilon(a,r)\) into mass grid bins at various radii.
\end{description}

Appendix \ref{appdenix:low_epsilon_approxi} provides an empirical fits for cases when the dust-to-gas ratio is too low to satisfy the criterion Eq.~\eqref{drift_criteria}. Utilizing the 2D size-spatial distribution for all dust species, we can reconstruct the dust ring for observational modeling through radiative transfer calculations across various (sub-)mm bands, which will be detailed in our upcoming paper.


\section{Conclusions} \label{sec:summary}

High-resolution, multi-wavelength observations have revealed ringed sub-structures in PPDs, typically indicating multi-species dust with size evolution. This observation necessitates self-consistent modeling that incorporates dust coagulation and fragmentation due to collisions among these species. Such modeling, however, often requires computationally expensive numerical simulations, which can limit the efficient extraction of key ring parameters, such as dust size and turbulence strength. To address this, we develop an analytical model for dust dynamics in these rings, calibrated using numerical simulations of dust coagulation in this study.

Starting from a gas pressure bump, we simulate dust-trapping rings with coagulation and fragmentation using \texttt{DustPy} \citep{Stammler_2022}. We find that in typical PPDs, coagulation and fragmentation dominate the dust evolution process, leading to a similar spatial distribution among different dust species in the coagulation-fragmentation equilibrium state. All species up to the fragmentation limit exhibit an intrinsic dust ring width \( w_{\epsilon} \) in the dust ring, as given by Eq.~(\ref{eq:wewg}).
This intrinsic width markedly differs from the drift-diffusion-only scenario commonly assumed in previous studies. 
Based on the size and spatial distribution of dust in the ring, we present a simplified model that accounts for all dust species. The application of our model, incorporating radiative transfer calculations at multiple wavelengths, will be presented in a separate paper to further constrain dust dynamics and coagulation in PPD rings.

\section*{Acknowledgements}
We thank the referee for many helpful suggestions that improve the quality of the paper. We also thank Chris Ormel, Shigeru Ida, Qiuyi Luo, and Bocheng Zhu for their instructive comments. We also acknowledge the Chinese Center for Advanced Science and Technology for hosting the Protoplanetary Disk and Planet Formation Summer School in 2022, organized by Xue-Ning Bai and Ruobing Dong, which inspired the ideas underlying this work.
L.Y and Y.P.L are supported in part by the Natural Science Foundation of China (grants 12373070, and 12192223), the Natural Science Foundation of Shanghai (grant NO. 23ZR1473700). H.B.L. is supported by the National Science and Technology Council (NSTC) of Taiwan (Grant Nos. 111-2112-M-110-022-MY3, 113-2112-M-110-022-MY3). The calculations have made use of the High Performance Computing Resource in the Core Facility for Advanced Research Computing at Shanghai Astronomical Observatory.

Softwares: \texttt{DustPy} \citep{Stammler_2022}, \texttt{Numpy} \citep{vanderWalt2011}, \texttt{Scipy} \citep{Virtanen2020}, \texttt{Matplotlib} \citep{Hunter2007}.

\begin{appendix}

\section{Equilibrium Dust Distribution}
\label{append:analytical_we}
The key physical processes around the dust ring are illustrated in Figure~\ref{fig:sketch}. Near the center of the pressure bump, both the diffusion and drift fluxes vanish. The dust distribution can be described by coagulation-fragmentation equilibrium analysis \citep{Birnstiel_2011}. The normalized dust size distribution profile depends on disk parameters and the coagulation model, expressed as
\begin{equation}
    \label{eq:sigma_d}
    \frac{\sigma_{\rm d}}{\Sigma_{\rm d}} \equiv \frac{1}{\Sigma_{\rm d}} \frac{\partial \Sigma_{\rm d}}{\partial\ln{a}} = {\rm func}(a, T, \Sigma_{\rm g}, \alpha, \rho_{\rm s}, v_{\rm frag}, \Omega_{\rm K}, \xi),
\end{equation}
where $\Sigma_{\rm d}$ is the total dust surface density and $\sigma_{\rm d}$ represents the dust density in logarithmic dust mass bins. The profile depends on the temperature $T$, gas density and turbulence strength $\alpha$, internal dust density $\rho_{\rm s}$, fragmentation velocity $v_{\rm frag}$, the local Keplerian frequency $\Omega_{\rm K}$, and the power-law index $\xi$ of the mass distribution of fragments, but is independent of the dust-to-gas ratio. 

The slope of the dust size distribution profile is determined by the collision mechanism and the fragment distribution, $\xi$. For small grains, the relative velocity may be dominated by Brownian motion rather than turbulence, resulting in a steeper slope. As $\xi$ increases, more mass is redistributed into smaller fragments after collisions, leading to a slope akin to a growth cascade.

The coagulation model is valid for cases where the fragmentation barrier, \(\mathrm{St_{frag}}\), is less than unity. Otherwise the highest collision velocity between grains is unable to reach the fragmentation velocity. Dust will grow uncontrollably, causing an unphysical depletion of dust mass in protoplanetary disks.

When the dust-to-gas ratio is sufficiently high, collisions become so efficient that the effects of drift and diffusion are negligible, i.e., \(\tau_{\mathrm{growth}} < \tau_{\mathrm{drift}}\). And the maximum dust size is determined by the fragmentation limit. In such cases, the equilibrium size distribution predicted by Eq.~\eqref{eq:sigma_d} is maintained at every radius in the pressure bump, while the total dust-to-gas ratio is adjusted by drift and diffusion.

Considering the mass conservation equation for each dust species \citep{Clarke1988},
\begin{equation}
    \label{eq:mass conserv}
    \frac{\partial \sigma_{\rm d}}{\partial t} + \frac{1}{r} \frac{\partial}{\partial r}\left\{r\left[ v_{\rm d} \sigma_{\rm d} - D_{\rm p} \Sigma_{\rm g} \frac{\partial}{\partial r} \left( \frac{\sigma_{\rm d}}{\Sigma_{\rm g}} \right) \right]\right\} = S_{\rm ext}(a).
\end{equation}

At a specified radius from the bump center, small grains are replenished by the diffusion flux from the central bump, while large grains are depleted due to drift toward the center (left-hand side).

There are several effects associated with coagulation. First, within a much shorter time interval ($\gtrsim \tau_{\rm growth}$, but $\ll \tau_{\rm diff},\tau_{\rm drift}$), the replenished small grains grow, and large grains fragment, leading to local mass redistribution, which is considered as the source term (right-hand side). The total source across different species balances out, as in steady state, the gain of one dust species exactly matches the loss of others. This means
\begin{equation}
    \int_{a_0}^{\infty} S\mathrm{_{ext}}(a) \,d\ln{a} = 0.
\end{equation}

Second, the slope and certain critical dust sizes -- such as the maximum grain size and the transition between two turbulent regimes, marked by the zig-zag features around \( 0.003\ {\rm cm} \) in Figure~\ref{fig:dust_distribution} -- remain unchanged with respect to radius, as confirmed in Figure~\ref{fig:dust_distribution}, i.e.,
\begin{equation}
\label{eq:const_B11}
    \frac{\partial}{\partial r}  \Big(\frac{\sigma_{\rm d}}{\Sigma_{\rm d}} \Big)= 0.
\end{equation}

After integrating over all dust species and simplifying mathematically, the integral of the mass conservation equation \eqref{eq:mass conserv} in steady state, where $\partial \sigma_{\rm d}/\partial t = 0$, can be expressed as:
\begin{equation}
    \label{eq:smplified mass conserv}
    \epsilon_{\rm{tot}} \int_{a_0}^{\infty}  v_{\rm d}\frac{\sigma_{\rm d}}{\Sigma_{\rm d}} \,d\ln{a}
    = 
    \frac{\partial \epsilon_{\rm{tot}}}{\partial r}\int_{a_0}^{\infty}D_{\rm p} \frac{\sigma_{\rm d}}{\Sigma_{\rm d}}\,d\ln{a}.
\end{equation}
Eq. (\ref{eq:smplified mass conserv}) can be solved for the dust-to-gas ratio using any gas disk profile. In this work, we focus on the case of a pure Gaussian gas profile, as defined in Eq.~(\ref{eq:P_bump}).

We further assume that the dependence of the Stokes number on radius is negligible:
\begin{equation}
    \frac{\partial}{\partial r} \mathrm{St} = 0.
\end{equation}
This assumption holds when the turbulence-driven fragmentation barrier dominates the size distribution (see Eq.~\ref{eq:st_frag}). Using the expressions for $v_{\rm d}$ (Eq.~\ref{eq:v_drift}) and $D_{\rm p}$ (Eq.~\ref{eq:Dp}), we obtain a simple solution of the dust ring width for the Gaussian bump (defined in Eq.~\ref{eq:epsilon}) in the form of Eq.~\eqref{eq:smplified mass conserv}:
\begin{equation}
    \label{eq:wewg0}
    w_{\epsilon} = w_{\rm p} \cdot \sqrt{\frac{\bar{D}}{\bar{V}}},
\end{equation}
where $\bar{V}$ and $\bar{D}$ are defined in Eqs.~(\ref{eq:vbar}, \ref{eq:dbar}).
A universal constant of 1.2 accounts for oversimplifications in the analytical derivation, such as assuming a constant maximum Stokes number across the bump. Importantly, $w_{\epsilon}$ remains independent of $\Sigma_{\rm g}$ and $\rho_{\rm s}$, marking a key distinction from the drift-diffusion balance solution. However, when dust mass predominantly accumulates in larger grains, the solution asymptotically approaches the single-species case: $w_{\epsilon} = w_{\rm p} \sqrt{\alpha/{\rm St}_{\rm max}}$, aligning with previous theoretical expectations.

In the presence of significant dust trapping, the steady-state dust-to-gas ratio at the ring center, $\epsilon_{\rm tot}(r_0)$, can be estimated as the initial dust-to-gas ratio: $\epsilon_{\rm tot}(r_0) \cdot w_{\epsilon} = \epsilon_{\rm ini} \cdot w_{\rm p}$, consistent with mass conservation in the bump region where $\Delta r < w_{\rm p} \sqrt{2 \ln{A_{\rm p}}}$.

\subsection{Criteria For Validity}
\begin{figure}
\includegraphics[width=\columnwidth]{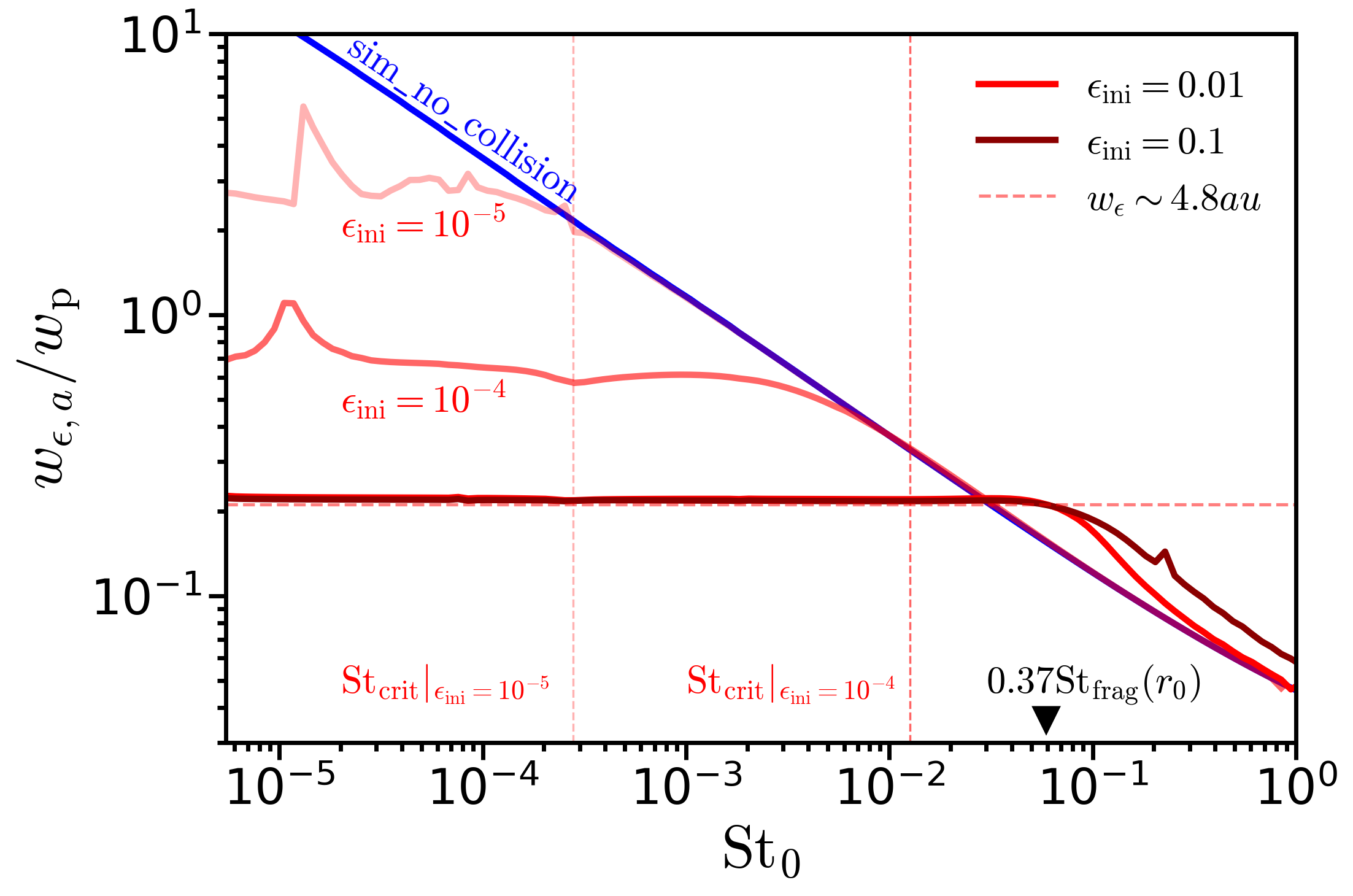}
\caption{Similar to Figure~\ref{fig:every_species}, this figure presents results for varying initial dust-to-gas ratios in the fiducial simulation. When $\epsilon_{\rm ini}$ is sufficiently high, dust grains below the fragmentation limit exhibit a common radial distribution, as discussed earlier. However, if $\epsilon_{\rm ini}$ is too low to satisfy the criteria in Eqs.~\eqref{drift_criteria} or \eqref{diff_criteria}, the dust population splits into two regimes, delineated by the critical Stokes number (Eq.~\ref{eq:St_crit}). Larger grains follow the drift-diffusion balance, while smaller grains maintain a common radial distribution.}
\label{fig:eps_series}
\end{figure}

Strictly speaking, coagulation-fragmentation equilibrium cannot be achieved if drift and diffusion are actively removing or replenishing dust. 
Quantitatively, the analytical model is valid when the dust growth timescale is shorter than the drift and diffusion timescales. By comparing growth timescales to drift timescales for grains with the size of $a_{\rm frag}$, we derive a critical minimum threshold for steady-state (rather than the initial one) dust-to-gas ratio at the bump center:
\begin{equation}
    \label{drift_criteria}
    \epsilon_{\rm tot,drift}(r_0) > \frac{v_{\rm frag}^2}{6 \alpha \Omega_{\rm K}^2 w_{\rm p}^2},
\end{equation}
which is equivalent to:
\begin{equation}
    \label{drift_criteria_wg}
    w_{\rm p} > \left( \sqrt{\frac{\mathrm{St_{frag}}}{2\epsilon_{\rm tot}}}H_{\mathrm{p}} \right) \Bigg|_{r_0},
\end{equation}
to ensure growth dominates over drift. For our fiducial simulation, this corresponds to an initial dust-to-gas ratio of $\epsilon_{\rm ini} \sim 1 \times 10^{-3}$. If this threshold is not met, drift instead of growth effects dominate, resulting in a significantly broader ring, as illustrated by semi-transparent red lines in Figure~\ref{fig:eps_series}.  Similarly, for diffusion effects, a minimum ratio is derived:
\begin{equation}
    \label{diff_criteria}
    \epsilon_{\rm tot,diff}(r_0) > \frac{D_{\rm p}}{w_{\rm p}^2 \Omega_{\rm K}} = \frac{\alpha c_{\rm s}^2}{w_{\rm p}^2 \Omega_{\rm K}^2 },
\end{equation}
below which diffusion dominates over growth for well-coupled dust particles. In our fiducial simulation, this threshold corresponds to an initial dust-to-gas ratio of $\epsilon_{\rm ini} \sim 1 \times 10^{-5}$. If the ratio drops below this threshold, dust particles hardly grow beyond micrometer sizes due to strong turbulence continually dispersing material.

\subsection{Low $\epsilon_{\rm tot}$ Approximation}
\label{appdenix:low_epsilon_approxi}
When the dust-to-gas ratio is very low, meaning the criterion in Eq. \eqref{drift_criteria} is not satisfied while Eq. \eqref{diff_criteria} holds, the dust size distribution at the bump center, \(\sigma_{\rm d}(r_0)\), can still be approximated by coagulation-fragmentation equilibrium. Here, we provide an empirical fit for the width. By comparing the growth timescale Eq.~\eqref{time:growth} with the drift timescale Eq.~\eqref{time:drift}, a critical Stokes number can be determined:
\begin{equation}
    \label{eq:St_crit}
    \mathrm{St_{crit}} = 2\epsilon_{\rm tot} \left(\frac{\Omega_{\rm K} w_{\rm p}}{c_{\rm s}}\right)^2.
\end{equation}
This threshold divides the dust ring into two distinct size regimes, as indicated by the vertical red dashed lines in Figure~\ref{fig:eps_series}:
\begin{description}
    \item[Small grains (\(\mathrm{St} < \mathrm{St_{crit}}/f\))] For these particles, the drift timescale remains longer than the growth timescale, and their radial distribution can be approximated as:
\begin{equation}
    w_{\epsilon,a_{\rm{small}}} \sim  w_{\rm p} \sqrt{f\frac{\alpha}{\mathrm{St_{crit}}}}.
\end{equation}
  \item[Large grains (\(\mathrm{St} > \mathrm{St_{crit}}/f\))] These grains are dominated by drift and follow the drift-diffusion balance:
\begin{equation}
    w_{\epsilon,a_{\rm{large}}} \sim w_{\rm p} \sqrt{\alpha/\mathrm{St_{0}}},
\end{equation}
\end{description}
where \(f\sim3\) is a numerical factor.

If the dust-to-gas ratio is extremely low, such that Eq.(\ref{diff_criteria}) is not satisfied, dust growth is inhibited by strong diffusion flux. However, this scenario is rare in typical protoplanetary disks and is therefore not explored in detail in this study.

\section{the Negligibility of Drift-Induced Fragmentation}

This study primarily focuses on turbulence-induced fragmentation. Here we demonstrate that fragmentation driven by relative drift can be safely ignored under typical conditions.

Dust grains experience differential coupling with the gas, leading to variations in drift, settling, and responses to turbulent stirring. As a result, grains exhibit relative motion and undergo collisions. In protoplanetary disks, two primary mechanisms contribute to the relative collision velocity: turbulence and drift.

The relative velocity due to turbulence is given by \citet{Ormel2007}:
\begin{equation}
\Delta v_{\rm turb} =\sqrt{3 \rm{St}}V_{\rm turb},
\end{equation}
where \(V_{\rm turb}\) is the characteristic turbulent velocity. In hydrodynamic turbulence, this velocity is approximated as \(V_{\rm turb}=\sqrt{\alpha}c_{\rm s}\) \citep{Youdin2007,Cui_Bai_2022}. Since \(\Delta v_{\rm turb}\) increases with the Stokes number, the maximum turbulence-driven relative velocity is:
\begin{equation}
\label{eq:turb_max}
\Delta v_{\rm turb,max} =\sqrt{3 \alpha \rm{St_{\rm max}}}c_{\rm s},
\end{equation}
where \(\rm{St_{\rm max}}\) represents the largest Stokes number attained by dust grains within the pressure bump.

In contrast, the relative drift velocity reaches a maximum between the largest and smallest grains within a Gaussian pressure bump. According to Eq.~\eqref{eq:v_drift}, the drift velocity is given by:
\begin{equation}
v_{\rm drift} = {{\rm St}}\frac{c_{\rm s}^2}{v_{\rm K}} \frac{r(r_0-r)}{{w_{\rm p}}^2},
\end{equation}
where the velocity depends on radial location, increasing towards the bump wings. The farthest radial location that can be reached by the largest grains is approximately \(r_0 - r \approx w_{\sigma_{\rm d},a} \sim w_{\rm p}\sqrt{\alpha/{\rm St_{max}}}\). Consequently, the maximum relative drift velocity within the bump is:
\begin{equation}
\label{eq:drift_max}
    \Delta v_{\rm drift,max} = {{\rm St_{max}}}\frac{c_{\rm s}^2}{v_{\rm K}} \frac{r(r_0-r)}{{w_{\rm p}}^2} \Bigg|_{r=r_0-w_{\rm p}\sqrt{\alpha/{\rm St_{max}}}}.
\end{equation}

By comparing Eqs. \eqref{eq:turb_max} and \eqref{eq:drift_max} the criterion for drift-induced fragmentation to dominate over turbulence-induced fragmentation (\(\Delta v_{\rm turb,max} < \Delta v_{\rm drift,max}\)) is:
\begin{equation}
    w_{\rm p} \lesssim H_{\rm p}/\sqrt{3},
\end{equation}
where \(H_{\rm p}\) is the gas scale height. Narrow bumps that satisfy this condition are more susceptible to drift-induced fragmentation. However, such rings are likely to trigger the RWI as discussed in Appendix~\ref{app:rwi} \citep{Eonho_Chang2023}, leading to crescent-shaped structures rather than axisymmetric rings. These effects are beyond the scope of this study.

\section{Are the Pressure Bump Rossby-Wave Unstable} \label{app:rwi}
In this work, we focus on the dust ring associated with a pressure bump.
While our analysis employs a one-dimensional axisymmetric model for the dust ring, it is well-established that pressure bumps can trigger non-axisymmetric structures through the RWI. This occurs because the strong pressure gradient creates a local vortensity extrema \citep{Lovelace_1999,Li2000,Ono_2016,Eonho_Chang2023}, as demonstrated in multi-dimensional simulations \citep[e.g.,][]{Li2001,Val-Borro_2007,Meheut_2010,Li2020}.

There exists a few linear analysis for the RWI, even though a general criterion for the onset of RWI may still lack \citep{Li2000,Ono_2016,Liu2023,Cui_2025}. 
The Lovelace criterion states that a necessary condition for the RWI is the presence of a local extrema in the vortensity profile of isentropic disks \citep{Lovelace_1999}. 
For Gaussian bumps, \citet{Ono_2016}  derived a necessary and sufficient condition for the onset of the RWI in semi-analytic form in a barotropic 2D disk.

\cite{Eonho_Chang2023} focused on the the role of RWI in dust trapping by performing a linear analysis of nonbarotropic disks, exploring a broader range of background disk slopes and comparing surface density bumps to temperature bumps. 
They suggested that the longest wavelength mode, $m=1$, corresponding to a single large-scale vortex, is usually the dominant mode of RWI.

The analyses are further complicated by other effects like dusts \citep{Kevin_2024,Liu2023,Cui_2025}, disk cooling \citep{Huang&Yu_2022,Fung_Ono_2021}, and non-ideal MHD \citep{Cui_2024,Cui_2025_b}.

To estimate the stability of the gas bump in this work, we check the stability boundary for $m=1$ modes of inviscid incompressible disk bumps given by \cite{Eonho_Chang_2024} (see Section 3.1 therein), which is an extension of the criterion from \citet{Ono_2016} and should be regarded as a conservative treatment. The stability boundary has two branches, depending a critical parameter:
\begin{equation}
  \mathcal{J}
    = \ln A_{\mathrm{p}}\,
      \frac{c_{\mathrm{s}}^2}{\gamma\,\Omega_{\mathrm{K}}^2\,w_{\mathrm{p}}^2}
    \Bigl\lvert_{r=r_0}\Bigr., 
\end{equation}
\medskip

Where $\gamma$ is the adiabatic index. If $\mathcal{J}\gg1$, the stable boundary is $w_{\rm _p}/r_0\gtrsim\sqrt{2}$, which suggests that wider bumps tend to be Rossby-wave stable. When $\mathcal{J}\lesssim1$, the stable condition is
\begin{equation}
    A_{\rm p}<1+1.2\left(\frac{w_{\rm _p}}{r_0}\right)^3\left(\frac{H_{\rm p,0}}{r_0}\right)^{-2}.
\end{equation}
Note that our definition of $A_{\rm p}$ differs from previous works by an offset of 1.0 (a unity background). For our fiducial model ($H_{\rm p,0}/r_{0}=0.05$, $w_{\rm p}/r_{0}=0.23$, $A_{\rm p}=2$), this ensures that the pressure bump is stable against RWI. While parameter surveys with disk temperature $T_0 \gtrsim 70\ \mathrm{K}$, which corresponds to a disk aspect ratio of $h\gtrsim 0.13$, may yield RWI-unstable configurations. Such a thick and hot disk at 100 au could be observationally inappropriate. Nevertheless, these results do not affect the conclusions of this work. 

\begin{figure}
\includegraphics[width=\columnwidth]{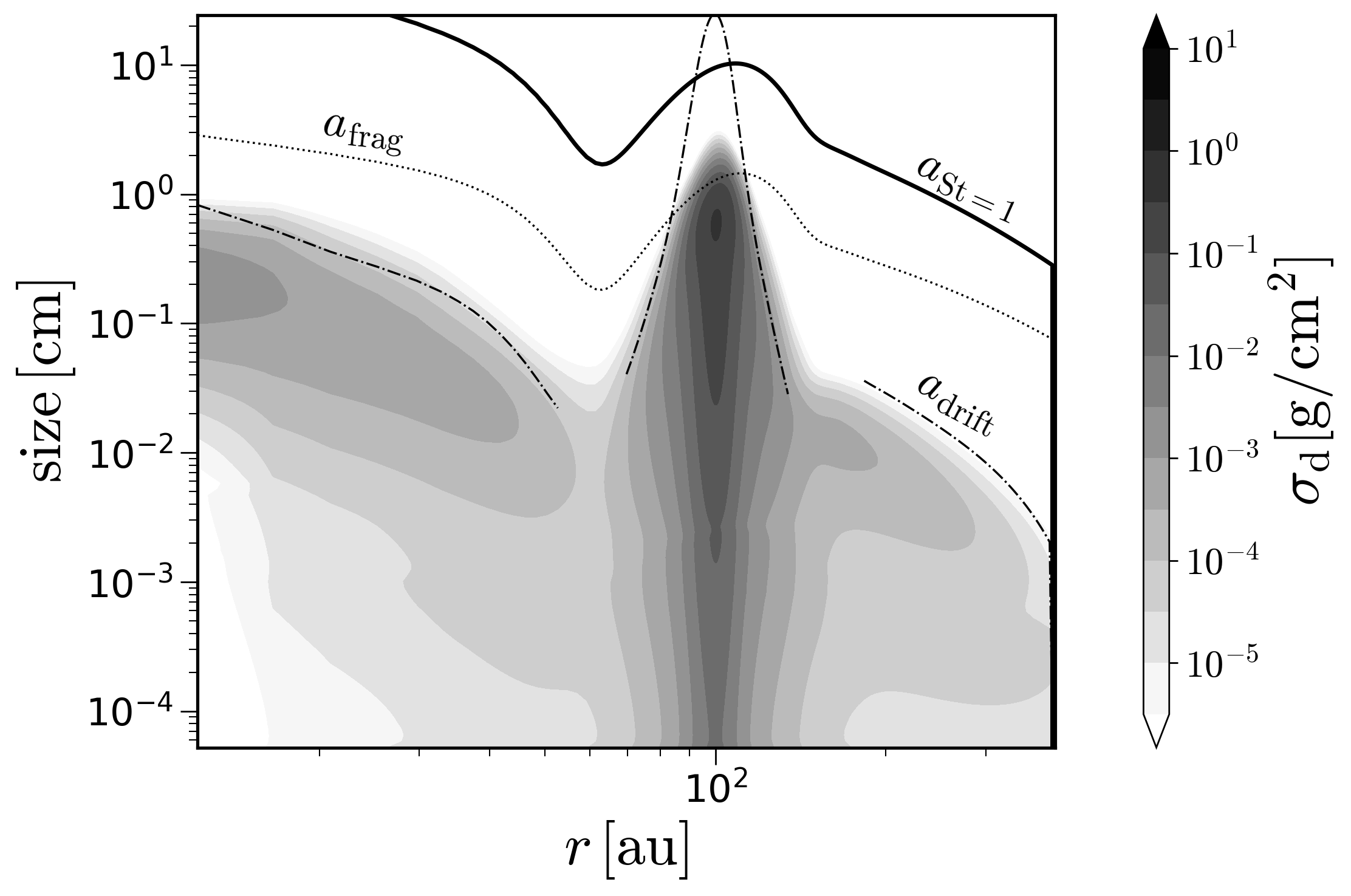}
\includegraphics[width=\columnwidth]{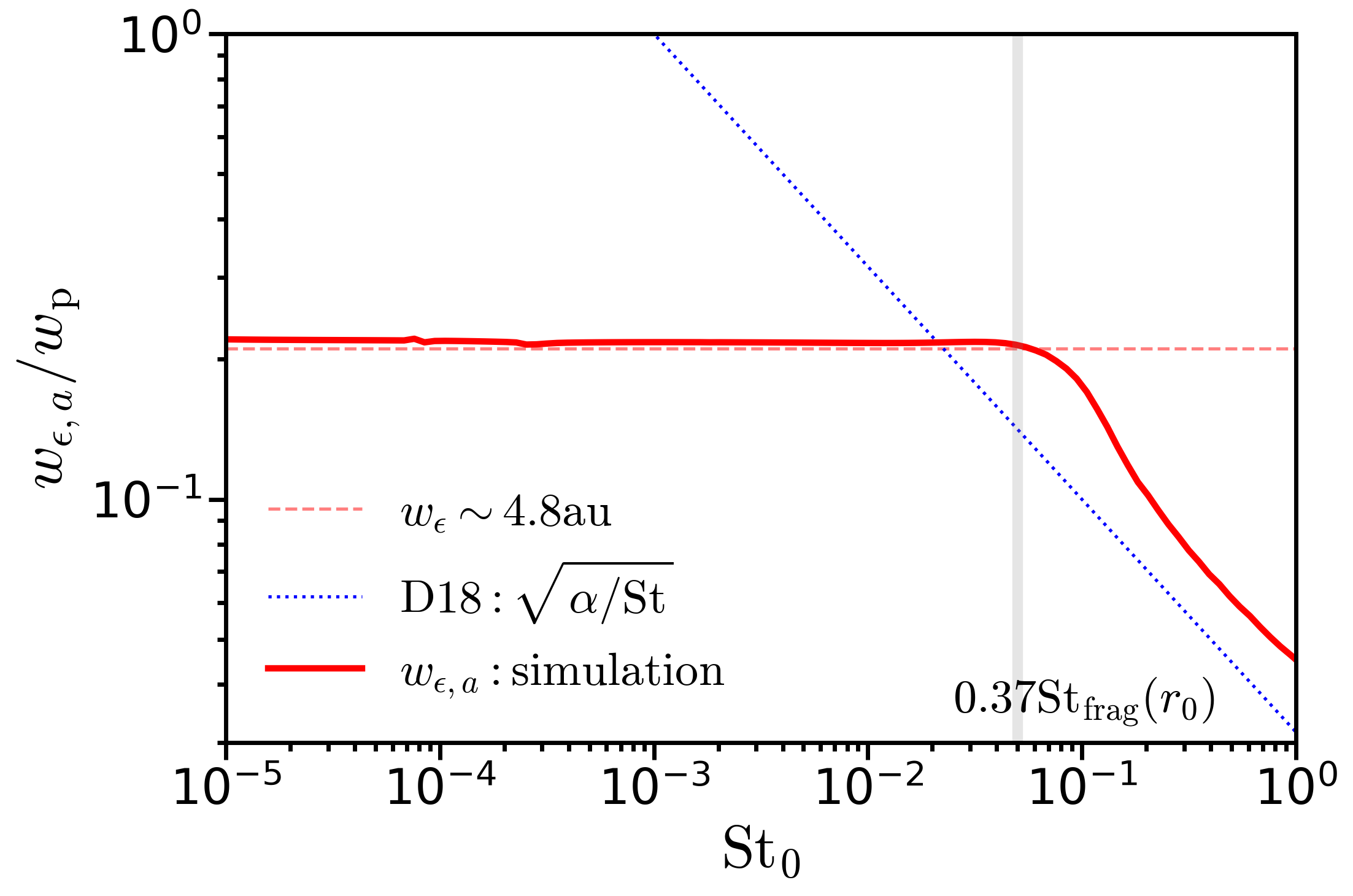}
\caption{Upper panel: Similar to Figure~\ref{fig:fiducial} except that \(A_{\rm p}=1.3\) for this case. Lower panel: Similar to Figure~\ref{fig:every_species}, but corresponding to the shallower bump as illustrated in the upper panel.}
\label{fig:shallow_width}
\end{figure}

To further illustrate the dependence of dust ring dynamics on the pressure bump, we simulate a shallower bump ($A_{\rm p}=1.3$) which corresponds to a more Rossby-wave stable configuration. The resulting dust size distribution and the ring width (Figure~\ref{fig:shallow_width})  match closely those of $A_{\rm p}=2$ in Figure~\ref{fig:every_species}. This demonstrates the robustness of our dust ring model.

We have further tested a non-Gaussian pressure bump profile from \citet{Eonho_Chang2023}, selecting parameters for marginal dust trapping and stable against the RWI (e.g., $A_{\Sigma}\simeq1.2$, $W/H_{0}\simeq4$; their Equation 1). Despite significant profile deformation, our dust ring model remains highly consistent with our simulation results.

\end{appendix}


\bibliography{citation.bib}{}
\bibliographystyle{aasjournalnolink}

\listofchanges  
\end{CJK*}
\end{document}